\documentclass[prd,preprintnumbers,amsmath,amssymb,nofootinbib]{revtex4}
\usepackage{graphics,color,subfigure}
\usepackage{epsfig}
\usepackage{dcolumn}% Align table columns on decimal point
\usepackage{bm}% bold math
\usepackage{longtable}

\usepackage[colorlinks=true,
                  urlcolor=blue,
                  linkcolor=blue,
                  citecolor=blue,
                  bookmarks=true,
                  bookmarksopen=true,
                  bookmarksnumbered=true]{hyperref}

\begin{document}
%\begin{CJK}{GBK}{song}
\title{Hidden-charm pentaquark states in a mass splitting model}
\author{Shi-Yuan Li$^1$}
\author{Yan-Rui Liu$^1$}\email{yrliu@sdu.edu.cn}
\author{Zi-Long Man$^1$}\email{manzilong@mail.sdu.edu.cn}
\author{Zong-Guo Si$^1$}
\author{Jing Wu$^2$}\email{wujing18@sdjzu.edu.cn}

\affiliation{$^1$School of Physics, Shandong University, Jinan, Shandong 250100, China\\
$^2$School of Science, Shandong Jianzhu University, Jinan 250101, China
}

\date{\today}

\begin{abstract}
Assuming that the $P^N_{\psi}(4312)^+$ is a $I(J^P)=\frac12(\frac32^-)$ compact pentaquark, we study the mass spectrum of its S-wave hidden-charm partner states in a color-magnetic interaction model. Combining the information from their decays obtained in a simple rearrangement scheme, one finds that the quantum numbers of $P^N_{\psi}(4457)^+$, $ P^N_{\psi}(4440)^+$, and $P^N_{\psi}(4337)^+$ can be assigned to be $I(J^P)=\frac12(\frac32^-)$, $\frac12(\frac12^-)$, and $\frac12(\frac12^-)$, respectively, while both $P^{\Lambda}_{\psi s}(4338)^0$ and $P^{\Lambda}_{\psi s}(4459)^0$ can be interpreted as $I(J^P)=0(\frac12^-)$ $udsc\bar{c}$ compact states. Based on the numerical results, we also find narrow pentaquarks in $ssnc\bar{c}$ ($n=u,d$) and $sssc\bar{c}$ systems. The decay properties of the studied pentaquarks and the searching channels for them can be tested in future experiments.
\end{abstract}

%\pacs{xxx}

%\end{CJK}
\maketitle

%%%%%%%%%%%%%%%%%%%%%%%%%%%%%%%%%%%%%%%%%%%
\section{Introduction}\label{sec1}
%%%%%%%%%%%%%%%%%%%%%%%%%%%%%%%%%%%%%%%%%%%

In 2015, two exotic states $P^N_{\psi}(4380)^+$ and $P^N_{\psi}(4450)^+$ were observed in the $J/\psi p$ invariant mass distributions in the decay $\Lambda_b^0 \to J/\psi pK^-$ by the LHCb Collaboration \cite{LHCb:2015yax}. Because their masses are very high, one cannot interpret them as excited three-quark baryons. Their minimal quark content should be $uudc\bar{c}$. Therefore, they are good candidates of hidden-charm pentaquark states. In 2019, LHCb\cite{LHCb:2019kea} reported a new hidden-charm pentaquark-like state $P^N_{\psi}(4312)^+$, while the $P^N_{\psi}(4450)^+$ were resolved into two states $P^N_{\psi}(4440)^+$ and $P^N_{\psi}(4457)^+$ with updated statistics. Recently, LHCb announced the evidence of new pentaquark-like states $P^N_{\psi}(4337)^+$, $P^{\Lambda}_{\psi s}(4459)^0$, and $P^{\Lambda}_{\psi s}(4338)^0$ in the decay channels $B^0_s \to J/\psi p\bar{p}$ \cite{LHCb:2021chn}, $\Xi_b^- \to J/\psi \Lambda K^-$ \cite{LHCb:2020jpq}, and $B^-\to J/\psi \Lambda \bar{p}$ \cite{LHCb:2023cph}, respectively. The minimal quark content of the $P^{\Lambda}_{\psi s}(4459)^0$ and $P^{\Lambda}_{\psi s}(4338)^0$ is $udsc\bar{c}$. We summarize the masses, decay widths, and observed channels of these states in Table \ref{experiment}. The newly observed hidden-charm pentaquark-like states inspired lot of debates about their inner structures and quantum numbers \cite{Ali:2019npk,Zhu:2019iwm,Giron:2019bcs,Stancu:2019qga,Giron:2021fnl,Deng:2022vkv,Wang:2019got,Wang:2020eep,Chen:2019asm,Li:2023zag,Chen:2019bip,Liu:2019tjn,Guo:2019kdc,Wang:2019ato,He:2019ify,Du:2019pij,Xiao:2019aya,Yamaguchi:2019seo,Wang:2022gfb,Du:2021fmf,Feijoo:2022rxf,Liu:2023wfo,Du:2021bgb,Peng:2020hql,Chen:2020uif,Chen:2020kco,Hu:2021nvs, Xiao:2021rgp,Zhu:2022wpi,Wang:2022nqs,Nakamura:2022gtu,Giachino:2022pws,Nakamura:2021dix,Nakamura:2021qvy,Park:2022nza}.

\begin{table}[htbp]\centering
	\caption{The masses, decay widths, and observed channels of the hidden-charm pentaquark-like states reported by the LHCb Collaboration.}\label{experiment}
	\begin{tabular}{cccc}\hline\hline
		& Mass(MeV)                          &$\Gamma$ (MeV)                   & observed channels \\ \hline
		$P^N_{\psi}(4380)^{+}$\cite{LHCb:2015yax}      &$ 4380\pm8\pm29$               &$215\pm18\pm86$            &$\Lambda_b^0 \to J/\psi pK^-$\\
		$P^N_{\psi}(4312)^+$\cite{LHCb:2019kea}        &$ 4311.9\pm0.7^{+6.8}_{-0.6}$   &$9.8\pm2.7^{+3.7}_{-4.5}$  &$\Lambda_b^0 \to J/\psi pK^-$\\
		$P^N_{\psi}(4440)^+$\cite{LHCb:2019kea}       &$ 4440\pm1.3^{+4.1}_{-4.7}$     &$20.6\pm4.9^{+8.7}_{-10.2}$  &$\Lambda_b^0 \to J/\psi pK^-$\\
		$P^N_{\psi}(4457)^+$\cite{LHCb:2019kea}       &$ 4457.3\pm0.6^{+4.1}_{-1.7}$   &$6.4\pm2.0^{+5.7}_{-1.9}$    &$\Lambda_b^0 \to J/\psi pK^-$\\
		$P^{\Lambda}_{\psi s}4459)^0$\cite{LHCb:2020jpq}  &$ 4458.8\pm2.9^{+4.7}_{-1.1}$  &$17.3\pm6.5^{+8.0}_{-5.7}$  &$\Xi_b^- \to J/\psi \Lambda K^-$\\
		$P^N_{\psi}(4337)^+$\cite{LHCb:2021chn}  &$ 4337^{+7\;+2}_{-4\;-2}$       &$29^{+26\;+14}_{-12\;-14}$  &$B^0_s \to J/\psi p\bar{p}$\\
		$P^{\Lambda}_{\psi s}(4338)^0$ \cite{LHCb:2023cph} &$ 4338.2\pm0.7\pm0.4$          &$7.0\pm1.2\pm1.3$          &$B^-\to J/\psi \Lambda \bar{p}$\\
		\hline\hline
	\end{tabular}
\end{table}

In the literature, interpretations of the above mentioned exotic baryons include compact pentaquark states \cite{Ali:2019npk,Zhu:2019iwm,Giron:2019bcs,Stancu:2019qga,Giron:2021fnl,Deng:2022vkv,Wang:2019got,Wang:2020eep},
molecule states \cite{Chen:2019asm,Li:2023zag,Chen:2019bip,Liu:2019tjn,Guo:2019kdc,Wang:2019ato,He:2019ify,Du:2019pij,Xiao:2019aya,Yamaguchi:2019seo,Du:2021fmf,Feijoo:2022rxf,Liu:2023wfo,Du:2021bgb,Peng:2020hql,Chen:2020uif,Chen:2020kco,Hu:2021nvs,Xiao:2021rgp,Zhu:2022wpi,Wang:2022nqs,Nakamura:2022gtu,Giachino:2022pws,Wang:2022gfb},
cusp effects \cite{Nakamura:2021dix,Nakamura:2021qvy},
coupled channel effects \cite{Yan:2021nio}, etc.
There are also studies of their decay and production properties \cite{Guo:2019fdo,Lin:2019qiv,Wang:2019spc,Gutsche:2019mkg,Burns:2019iih,Xiao:2020frg,Burns:2022uiv,Lin:2023dbp,Azizi:2021utt,Wu:2021caw}.
One may consult Refs. \cite{Liu:2019zoy,Chen:2022asf,Meng:2022ozq,Mutuk:2019snd} for more discussions. Most studies support the molecule interpretation.
In fact, one still hardly distinguishs the inner structures of these observed pentatquark-like states from the current experimental data. The possibility that their properties can be understood in the compact pentaquark picture is still not ruled out.

In previous papers \cite{Wu:2017weo,Cheng:2019obk}, we have studied the mass spectra and rearrangement decays of S-wave hidden-charm pentaquark states with the $(qqq)_{8_c}(Q\bar{Q})_{8_c}$ ($q=u,d,s$) configuration in the chromomagnetic interaction (CMI) model by choosing a reference hadron-hadron channel. From the combined analysis of spectrum and decay, our results indicate that the $P^N_{\psi}(4457)^+$, $P^N_{\psi}(4440)^+$, and $P^N_{\psi}(4312)^+$ are probably $J^P=3/2^-$, $1/2^-$, and $3/2^-$ $(uud)_{8_c}(c\bar{c})_{8_c}$ pentaquark states, respectively. However, there are two drawbacks in these works. On the one hand, the mass spectra are estimated by using a hadron-hadron threshold as the reference scale and the choice of meson-baryon channel affects the results. On the other hand, the contributions from the color-singlet $(qqq)_{1_c}(c\bar{c})_{1_c}$ component were not considered in the pentaquark wave functions, which caused the lack of information of charmonium decay channels. Here, we revisit the compact hidden-charm pentaquark states with an improved framework.

In Ref. \cite{Wu:2016gas}, we found that the $\chi_{c1}(4140)$ can be interpreted as a compact $cs\bar{c}\bar{s}$ tetraquark state. Later in Refs. \cite{Wu:2018xdi,Cheng:2020nho}, we found that the mass spectra of other tetraquarks may be obtained by treating the $\chi_{c1}(4140)$ as a reference state. Now, we use a similar idea to study compact pentaquarks. We improve the CMI model to estimate masses of the hidden-charm pentaquark states assuming that the $P^N_{\psi}(4312)^+$ is a compact pentaquark with $J^P=\frac32^-$. This assumption differs from that in the molecule picture where the $P^N_{\psi}(4312)^+$ is a $\Sigma_c\bar{D}$ molecule with $J^P=\frac12^-$. The main reason why we adopt this assumption is from the consideration on the theoretical side. Since the estimated masses for the compact pentaquarks \cite{Cheng:2019obk} have some uncertainties, the assignment for the spin of an observed state is not unique. We further tried to make a reasonable assignment for the observed states from their decay information. By exploring the width ratios between different pentaquarks with various assignments, we found that the $J^P=\frac32^-$ for $P^N_{\psi}(4312)^+$ correspond to the most appropriate assignment. Another reason for using the assumption is that the spin of $P^N_{\psi}(4312)^+$ has not been determined experimentally yet. We are going to investigate a different possibility from the molecule picture for the nature of the observed pentaquark-like exotic states in a self-consistent way.

Up to now, all the hidden-charm pentaquarks are observed in the $J/\psi$ channels. It is necessary to include the $(qqq)_{1_c}(c\bar{c})_{1_c}$ components in the wavefunctions of the compact pentaquark states. By comparing the theoretical calculations and experimental data, such decay properties can provide more information about the internal structures of hadrons. Therefore, we also include the hidden-charm channels in the calculation of decay widths with a simple scheme.

This paper is arranged as follows. After the introduction, we present the formalism to study mass spectra and rearrangement decays of hidden-charm pentaquark states in Sec. \ref{sec2}. The numerical results which include discussions about predicted stable pentaquarks and possible assignments for the observed states will be given in Sec. \ref{sec3}. The last section is for summary.

%%%%%%%%%%%%%%%%%%%%%%%%%%%%%%%%%%%%%%%%%%%
\section{Formalism }\label{sec2}
%%%%%%%%%%%%%%%%%%%%%%%%%%%%%%%%%%%%%%%%%%%

\subsection{Mass splitting model}
We employ the chromomagnetic interaction model to study the S-wave $qqqc\bar{c}$ ($q=u,d,s$) systems. The model Hamiltonian reads
\begin{equation}
H=\sum_{i}m_i+H_{CMI}=\sum_{i}m_i-\sum_{i<j}C_{ij}\lambda_{i}\cdot\lambda_j\sigma_i\cdot\sigma_j,
\end{equation}
where $\lambda_{i}$ and $\sigma_{i}$ are the Gell-Mann matrix and the Pauli matrix for the $i$-th quark, respectively. $m_i$ is effective quark mass. The effective coupling coefficient $C_{ij}$ reflects the strength between the $i$-th and $j$-th quarks, which can be extracted from the ground hadrons. One calculates the mass of an S-wave pentaquark with
\begin{equation}\label{mass1}
M=\sum_{i}m_i+\langle H_{CMI}\rangle
\end{equation}
after diagonalizing the Hamiltonian. In fact, we obtained overestimated hadron masses with this formula in our previous studies. They may be regarded as theoretical upper limits \cite{Wu:2016vtq,Luo:2017eub,Wu:2017weo,Wu:2016gas,Zhou:2018bkn,Liu:2019zoy,Li:2018vhp}. The overestimated masses are mainly due to the values of $m_i$'s. Because each system actually has its own $m_i's$, the model can not afford an appropriate description of the attraction between quark components for all systems. To get more reasonable theoretical results, the mass of a pentaquark state can be rewritten as
\begin{equation}
M=(M_{ref}-\langle H_{CMI}\rangle_{ref})+\langle H_{CMI}\rangle,
\end{equation}
where $M_{ref}$ and $\langle H_{CMI}\rangle_{ref}$ are the measureed mass and choromomagnetic interaction matrix of the reference system, respectively. This method can partially compensate the uncertainty caused by effective quark masses \cite{Wu:2018xdi}.

There are two schemes to choose the reference system for the hidden-charm pentaquark states. The first scheme involves a meson-baryon channel whose threshold is treated as the reference scale. It yields more reasonable results than the scheme adopting Eq. \eqref{mass1}. In our previous work \cite{Wu:2017weo}, we obtained masses of hidden-charm pentaquark states with different thresholds, but it is difficult to determine which threshold is a more appropriate choice. The second scheme adopts a compact reference pentaquark, which is more reasonable than the first scheme since the structure of a meson-baryon state is actually different from a compact state. The procedure is similar to getting the estimated masses for tetraquark systems, where one identifys the $\chi_{c1}(4140)$ as the lowest $1^{++}$ $cs\bar{c}\bar{s}$ compact tetraquark and treats it as the reference state \cite{Wu:2018xdi,Cheng:2020nho}. In Ref. \cite{Cheng:2019obk}, we studied the $(uud)_{8_c}(c\bar{c})_{8_c}$ pentaquark states within the CMI model using a (charmed meson)-(charmed baryon) threshold as a reference. The results indicate that the $P^N_{\psi}(4312)^+$ can be assigned as a $J=\frac32$ $(uud)_{8_c}(c\bar{c})_{8_c}$ compact pentaquark. Here, we still assume that the $P^N_{\psi}(4312)^+$ is a compact state with $I(J^P)=\frac12(\frac32^-)$ and choose it as a reference in the present case. The difference is that it is now a mixed state of $(uud)_{8_c}(c\bar{c})_{8_c}$ and $(uud)_{1_c}(c\bar{c})_{1_c}$. From the following numerical results (see Sec. \ref{sec.3.2}), one finds that the colored $(qqq)_{8_c}(c\bar{c})_{8_c}$ component of $P^N_{\psi}(4312)^+$ plays a dominant role in the wave function and the adopted assumption is consistent with Ref. \cite{Cheng:2019obk}. In this updated scheme, the mass formulas for the considered systems are
\begin{eqnarray}
&&M_{nnnc\bar{c}}=(M_{P^N_{\psi}(4312)^+}-\langle H_{CMI}\rangle_{P^N_{\psi}(4312)^+})+\langle H_{CMI}\rangle_{nnnc\bar{c}},\\
&&M_{nnsc\bar{c}}=(M_{P^N_{\psi}(4312)^+}-\langle H_{CMI}\rangle_{P^N_{\psi}(4312)^+})+\Delta_{sn}+\langle H_{CMI}\rangle_{nnsc\bar{c}}, \\
&&M_{ssnc\bar{c}}=(M_{P^N_{\psi}(4312)^+}-\langle H_{CMI}\rangle_{P^N_{\psi}(4312)^+})+2\Delta_{sn}+\langle H_{CMI}\rangle_{ssnc\bar{c}}, \\
&&M_{sssc\bar{c}}=(M_{P^N_{\psi}(4312)^+}-\langle H_{CMI}\rangle_{P^N_{\psi}(4312)^+})+3\Delta_{sn}+\langle H_{CMI}\rangle_{sssc\bar{c}},
\end{eqnarray}
where $\Delta_{sn}=m_s-m_n$ denotes the effective quark mass gap between $s$ quark and $n(=u,d)$ quark.
To relate the masses of $nnsc\bar{c}$, $nssc\bar{c}$, and $sssc\bar{c}$ to that of $P^N_{\psi}(4312)^+$, we introduce this parameter. Compared with Eq. \eqref{mass1}, the problem of effective quark mass becomes that of mass gap between different flavors of quarks and the uncertainty caused by effective quark masses are partially compensated \cite{Wu:2018xdi}.

To calculate the CMI Hamiltonians of pentaquark systems, one constructs their wave functions. In Refs. \cite{Wu:2017weo,Cheng:2019obk}, the wave functions involving color-octet component $(qqq)_{8_c}(c\bar{c})_{8_c}$ have been obtained. In the present work, we reconstruct wave functions by incorporating the color-singlet component $(qqq)_{1_c}(c\bar{c})_{1_c}$. These wave functions which are summarized in Table \ref{WF} will also be used to understand the decay properties of hidden-charm pentaquark states. In the table, we adopt the notation $[(qqq_{flavor})^{spin}_{color}(c\bar{c})^{spin}_{color}]^{spin}_{color}$. For brevity, we use $F$ ($D$) to denote the flavor wave function of the first three (two) light quarks. The notation $MS$ ($MA$) indicates that the first two light quarks are symmetric (antisymmetric) and $S$ ($A$) means that the wave function is totally symmetric (antisymmetric) in flavor, spin, or color space. For example, the wave function $[(F_S)^{S}_{A}(c\bar{c})^{1}_{1}]^{\frac52}_1$ is for the $I(J)=\frac32(\frac52)$ case. The subscript $S$ in $F_S$ indicates that the flavor wave function for the first three quarks is symmetric under the permutation of any two quarks and the superscript $S$ (subscript $A$) of $F_S$ means that the spin (color) wave function for the three light quarks is totally symmetric (antisymmetric).

\begin{table}[htbp]
	\caption{Possible wave functions for the hidden-charm pentaquark states with notation $[(qqq_{flavor})^{spin}_{color}(c\bar{c})^{spin}_{color}]^{spin}_{color}$ ($q=u,d,s$). $F$ ($D$) represents the flavor wave function of the first three (two) light quarks. $MS$ ($MA$) means that the wave function is symmetric (antisymmetric) under the permutation of the first two light quarks and $S$ ($A$) means that the wave function is symmetric (antisymmetric) under the permutation of any two light quarks. }\label{WF}
	\begin{tabular}{cl}\hline\hline
		$(I=\frac32,Y=1,J=\frac52)$&$[(F_S)^{S}_{A}(c\bar{c})^{1}_{1}]^{\frac52}_1$\\
		$(I=\frac32,Y=1,J=\frac32)$&$\frac{1}{\sqrt{2}}\{[(F_S)^{MS}_{MA}(c\bar{c})^1_8]^{\frac32}_1-[(F_S)^{MA}_{MS}(c\bar{c})^1_8]^{\frac32}_1\}$;
		$[(F_S)^S_{A}(c\bar{c})^{0}_{1}]^{\frac32}_1$;$[(F_S)^{S}_{A}(c\bar{c})^{1}_{1}]^{\frac32}_1$
		\\
		$(I=\frac32,Y=1,J=\frac12)$
		&$\frac{1}{\sqrt{2}}\{[(F_S)^{MS}_{MA}(c\bar{c})^0_8]^{\frac12}_1-[(F_S)^{MA}_{MS}(c\bar{c})^0_8]^{\frac12}_1\}$;
		$\frac{1}{\sqrt{2}}\{[(F_S)^{MS}_{MA}(c\bar{c})^1_8]^{\frac12}_1-[(F_S)^{MA}_{MS}(c\bar{c})^1_8]^{\frac12}_1\};$\\
		&$[(F_S)^{S}_{A}(c\bar{c})^{1}_{1}]^{\frac12}_1$\\ \hline
		
		$(I=\frac12,Y=1,J=\frac52)$
		&$\frac{1}{\sqrt{2}}\{[(F_{MS})^{S}_{MA}(c\bar{c})_8^1]^{\frac52}_{{1}}-[(F_{MA})^{S}_{MS}(c\bar{c})_8^1]^{\frac52}_{1}\}$\\
		$(I=\frac12,Y=1,J=\frac32)$
		&$\frac12\{ [(F_{MS})^{MA}_{MS}(c\bar{c})_8^1]^{\frac32}_{{1}}+[(F_{MA})^{MS}_{MS}(c\bar{c})_8^1]^{\frac32}_{{1}}+[(F_{MS})^{MS}_{MA}(c\bar{c})_8^1]^{\frac32}_{{1}}-[(F_{MA})^{MA}_{MA}(c\bar{c})_8^1]^{\frac32}_{{1}}\}$\\
		&$\frac{1}{\sqrt{2}}\{[(F_{MS})^{S}_{MA}(c\bar{c})_8^0]^{\frac32}_{{1}}-[(F_{MA})^{S}_{MS}(c\bar{c})_8^0]^{\frac32}_{{1}}\}$\\
		&$\frac{1}{\sqrt{2}}\{[(F_{MS})^{S}_{MA}(c\bar{c})_8^1]^{\frac32}_{{1}}-[(F_{MA})^{S}_{MS}(c\bar{c})_8^1]^{\frac32}\}_{{1}}$\\
		&$\frac{1}{\sqrt{2}}\{[(F_{MS})^{MS}_{A}(c\bar{c})_1^1]^{\frac32}_{{1}}+[(F_{MA})^{MA}_{A}(c\bar{c})_1^1]^{\frac32}\}_{{1}}$\\
		$(I=\frac12,Y=1,J=\frac12)$
		&$\frac12\{ [(F_{MS})^{MA}_{MS}(c\bar{c})_8^0]^{\frac12}_{{1}}+[(F_{MA})^{MS}_{MS}(c\bar{c})_8^0]^{\frac12}_{{1}}+[(F_{MS})^{MS}_{MA}(c\bar{c})_8^0]^{\frac12}_{{1}}-[(F_{MA})^{MA}_{MA}(c\bar{c})_8^0]^{\frac12}_{{1}}\}$\\
		&$\frac12\{ [(F_{MS})^{MA}_{MS}(c\bar{c})_8^1]^{\frac12}_{{1}}+[(F_{MA})^{MS}_{MS}(c\bar{c})_8^1]^{\frac12}_{{1}}+[(F_{MS})^{MS}_{MA}(c\bar{c})_8^1]^{\frac12}_{{1}}-[(F_{MA})^{MA}_{MA}(c\bar{c})_8^1]^{\frac12}_{{1}}\}$\\
		&$\frac{1}{\sqrt{2}}\{[(F_{MS})^{S}_{MA}(c\bar{c})_8^1]^{\frac12}_{{1}}-[(F_{MA})^{S}_{MS}(c\bar{c})_8^1]^{\frac12}_{{1}}\}$\\
		&$\frac{1}{\sqrt{2}}\{[(F_{MS})^{MS}_{A}(c\bar{c})_1^0]^{\frac12}_{{1}}+[(F_{MA})^{MA}_{A}(c\bar{c})_1^0]^{\frac12}_{{1}}\}$\\
		&$\frac{1}{\sqrt{2}}\{[(F_{MS})^{MS}_{A}(c\bar{c})_1^1]^{\frac12}_{{1}}+[(F_{MA})^{MA}_{A}(c\bar{c})_1^1]^{\frac12}_{{1}}\}$\\ \hline
		$(I=1,Y=0,J=\frac52)$&$[(D_Ss)^{S}_{MA}(c\bar{c})^1_8]_1^{\frac52};\;[(D_Ss)^{S}_{A}(c\bar{c})^1_1]_1^{\frac52}$\\
		
		$(I=1,Y=0,J=\frac32)$&$[(D_Ss)^{MS}_{MA}(c\bar{c})^1_8]_1^{\frac32};\; [(D_Ss)^{S}_{MA}(c\bar{c})^0_8]_1^{\frac32};\;[(D_Ss)^{S}_{MA}(c\bar{c})^1_8]_1^{\frac32}$;\\
		&$[(D_Ss)^{MA}_{MS}(c\bar{c})^1_8]_1^{\frac32};\; [(D_Ss)^{MS}_{A}(c\bar{c})^1_1]_1^{\frac32};\ [(D_Ss)^{S}_{A}(c\bar{c})^0_1]_1^{\frac32};\;[(D_Ss)^{S}_{A}(c\bar{c})^1_1]_1^{\frac32}$\\
		$(I=1,Y=0,J=\frac12)$& $[(D_Ss)^{MS}_{MA}(c\bar{c})^0_8]_1^{\frac12};\; [(D_Ss)^{MS}_{MA}(c\bar{c})^1_8]_1^{\frac12};\; [(D_Ss)^{S}_{MA}(c\bar{c})^1_8]_1^{\frac12};\;  [(D_Ss)^{MA}_{MS}(c\bar{c})^0_8]_1^{\frac12};$\\
		&$[(D_Ss)^{MA}_{MS}(c\bar{c})^1_8]_1^{\frac12};\;[(D_Ss)^{MS}_{A}(c\bar{c})^0_1]_1^{\frac12};\;[(D_Ss)^{MS}_{A}(c\bar{c})^1_1]_1^{\frac12};\;[(D_Ss)^{S}_{A}(c\bar{c})^1_1]_1^{\frac12} $ \\ \hline
		$ (I=0,Y=0,J=\frac52)$&$[(D_As)^{S}_{MS}(c\bar{c})^1_8]_1^{\frac52}$\\
		$(I=0,Y=0,J=\frac32)$&  $[(D_As)^{MS}_{MS}(c\bar{c})^1_8]_1^{\frac32};\;[(D_As)^{S}_{MS}(c\bar{c})^0_8]_1^{\frac32};\;[(D_As)^{S}_{MS}(c\bar{c})^1_8]_1^{\frac32};\;[(D_As)^{MA}_{MA}(c\bar{c})^1_8]_1^{\frac32};\;
		[(D_As)^{MA}_A(c\bar{c})^1_1]_1^{\frac32}$\\
		$ (I=0,Y=0,J=\frac12)$& $[(D_As)^{MS}_{MS}(c\bar{c})^0_8]_1^{\frac12};\;[(D_As)^{MS}_{MS}(c\bar{c})^1_8]_1^{\frac12};\;[(D_As)^{S}_{MS}(c\bar{c})^1_8]_1^{\frac12};\;[(D_As)^{MA}_{MA}(c\bar{c})^0_8]_1^{\frac12}\;[(D_As)^{MA}_{MA}(c\bar{c})^1_8]_1^{\frac12}\;$\\
		&$[(D_As)^{MA}_A(c\bar{c})^0_1]_1^{\frac12};\;[(D_As)^{MA}_A(c\bar{c})^1_1]_1^{\frac12}$\\ \hline
		
		$(I=\frac12,Y=-1,J=\frac52)$&$[(D_S{n})^{S}_{MA}(c\bar{c})^1_8]_1^{\frac52};\;[(D_S{n})^{S}_{A}(c\bar{c})^1_1]_1^{\frac52}$\\
		
		$(I=\frac12,Y=-1,J=\frac32)$&$[(D_S{n})^{MS}_{MA}(c\bar{c})^1_8]_1^{\frac32};\; [(D_S{n})^{S}_{MA}(c\bar{c})^0_8]_1^{\frac32};\;[(D_S{n})^{S}_{MA}(c\bar{c})^1_8]_1^{\frac32}$;\\
		&$[(D_S{n})^{MA}_{MS}(c\bar{c})^1_8]_1^{\frac32};\; [(D_S{n})^{MS}_{A}(c\bar{c})^1_1]_1^{\frac32};\ [(D_S{n})^{S}_{A}(c\bar{c})^0_1]_1^{\frac32};\;[(D_S{n})^{S}_{A}(c\bar{c})^1_1]_1^{\frac32}$\\
		$(I=\frac12,Y=-1,J=\frac12)$& $[(D_S{n})^{MS}_{MA}(c\bar{c})^0_8]_1^{\frac12};\; [(D_S{n})^{MS}_{MA}(c\bar{c})^1_8]_1^{\frac12};\; [(D_S{n})^{S}_{MA}(c\bar{c})^1_8]_1^{\frac12};\;  [(D_S{n})^{MA}_{MS}(c\bar{c})^0_8]_1^{\frac12};$\\
		&$[(D_S{n})^{MA}_{MS}(c\bar{c})^1_8]_1^{\frac12};\;[(D_S{n})^{MS}_{A}(c\bar{c})^0_1]_1^{\frac12};\;[(D_S{n})^{MS}_{A}(c\bar{c})^1_1]_1^{\frac12};\;[(D_S{n})^{S}_{A}(c\bar{c})^1_1]_1^{\frac12} $ \\ \hline
		
		$(I=0,Y=-2,J=\frac52)$&$[(F_S)^{S}_{A}(c\bar{c})^{1}_{1}]^{\frac52}_1$\\
		$(I=0,Y=-2,J=\frac32)$&$\frac{1}{\sqrt{2}}\{[(F_S)^{MS}_{MA}(c\bar{c})^1_8]^{\frac32}_1-[(F_S)^{MA}_{MS}(c\bar{c})^1_8]^{\frac32}_1\}$;
		$[(F_S)^S_{A}(c\bar{c})^{0}_{1}]^{\frac32}_1$;$[(F_S)^{S}_{A}(c\bar{c})^{1}_{1}]^{\frac32}_1$
		\\
		$(I=0,Y=-2,J=\frac12)$
		&$\frac{1}{\sqrt{2}}\{[(F_S)^{MS}_{MA}(c\bar{c})^0_8]^{\frac12}_1-[(F_S)^{MA}_{MS}(c\bar{c})^0_8]^{\frac12}_1\}$;
		$\frac{1}{\sqrt{2}}\{[(F_S)^{MS}_{MA}(c\bar{c})^1_8]^{\frac12}_1-[(F_S)^{MA}_{MS}(c\bar{c})^1_8]^{\frac12}_1\};$\\
		&$[(F_S)^{S}_{A}(c\bar{c})^{1}_{1}]^{\frac12}_1$\\
		\hline\hline
	\end{tabular}
\end{table}

Here, we present the calculated CMI matrices with explicit expressions. For the $I=\frac{3}{2},Y=1$ case, we have
\begin{gather}
\langle H_{CMI}\rangle_{J=\frac52}=8C_{12}+\frac{16}{3}C_{45};
\end{gather}
\begin{gather}
\langle H_{CMI}\rangle_{J=\frac32}=\begin{pmatrix}
10C_{12}+\frac{10}{3}C_{14}-\frac{10}{3}C_{15}-\frac23C_{45}  &\frac{8}{\sqrt{3}}(C_{14}+4C_{15})  &\frac{8\sqrt{5}}{3}(-C_{14}+C_{15})\\
&8(C_{12}-2C_{45})                   &0                       \\
&                                              &8C_{12}+\frac{16}{3}C_{45}                \\
\end{pmatrix};
\end{gather}
\begin{gather}
\langle H_{CMI}\rangle_{J=\frac12}=\begin{pmatrix}
10C_{12}+2C_{45}    &\frac{10}{\sqrt{3}}(C_{14}+C_{15})  &-8\sqrt{\frac{2}{3}}(C_{14}+C_{15})\\
&10C_{12}-\frac{20}{3}(C_{14}-C_{15})-\frac23C_{45}    &\frac{8\sqrt{2}}{3}(-C_{14}+C_{15})     \\
&                                              &8C_{12}+\frac{16}{3}C_{45}
\end{pmatrix}.
\end{gather}
For the $I=\frac{1}{2},Y=1$ case, the matrices are
\begin{gather}
\langle H_{CMI}\rangle_{J=\frac52}=2C_{12}+6C_{14}+6C_{15}-\frac23C_{45};
\end{gather}
\begin{gather}
\langle H\rangle_{J=\frac32}=\begin{pmatrix}
-2C_{12}+2C_{14}+2C_{15}-\frac23C_{45}  &2\sqrt{\frac{2}{3}}(C_{14}+4C_{15})           &-\frac23\sqrt{10}(C_{14}-4C_{15})   &\frac83(C_{14}-C_{15})\\
&2(C_{12}+C_{45})                              &2\sqrt{15}(C_{14}-C_{15})           &4\sqrt{\frac{2}{3}}(C_{14}+C_{15}) \\
&                                              &2C_{12}-\frac23(6c_{14}+6C_{15}+C_{45})&\frac43\sqrt{10}(-C_{14}+C_{15})                     \\
&                                   &                         &-8C_{12}+\frac{16}{3}C_{45}
\end{pmatrix};
\end{gather}
\begin{gather}
\langle H_{CMI}\rangle_{J=\frac12}=\begin{pmatrix}
-2C_{12}+2C_{45}  &2\sqrt3(C_{14}-C_{15})                     &-\frac{4}{\sqrt{3}}(C_{14}+4C_{15})  &0&\frac{8}{\sqrt{3}}(C_{14}+C_{15})\\
& -\frac23\left(\begin{array}{c}3C_{12}+6C_{14}\\+6C_{15}+C_{45}\end{array}\right)  &-\frac43(C_{14}-4C_{15}) &\frac{8}{\sqrt{3}}(C_{14}+C_{15})&-\frac{16}{3}(C_{14}-C_{15})\\
&                                      &\left(\begin{array}{c}2C_{12}-10C_{14}\\-10C_{15}-\frac23C_{45}\end{array}\right)&-\frac{8}{\sqrt{3}}(C_{14}+C_{15})&-\frac{8}{3}(C_{14}-C_{15})   \\
&                                   &                         &-8C_{12}-16C_{45} &0                 \\
&                                  &                         &                  &-8C_{12}+\frac{16}{3}C_{45}
\end{pmatrix}.
\end{gather}

Now we move on to the $nnsc\bar{c}$ systems. For simplicity, we write the CMI matrix in the form
\begin{gather}
\langle H_{CMI}\rangle=\begin{pmatrix}
X&Y\\
Y^T&Z
\end{pmatrix},
\end{gather}
where the symmetric matrix $X$ involves only color-octet contributions and the symmetric $Z$ is for the color-singlet component. The $X$ expressions can be found in Ref. \cite{Cheng:2019obk}. We here just give $Y$ and $Z$ results. For the $I=1,Y=0$ case, we have
\begin{gather}
Y_{J=\frac52}=\frac{4\sqrt{2}}{3}(\beta-\nu),
\quad
Z_{J=\frac52}=\frac83(C_{12}+2C_{13}+2C_{45});
\end{gather}
\begin{gather}
Y_{J=\frac32}=\begin{pmatrix}
\frac{4\sqrt{2}}{9}(2\beta+\nu)                     &\frac43\sqrt{\frac23}(\alpha+2\mu)             &-\frac{4\sqrt{10}}{9}(\beta+2\nu)\\
\frac43\sqrt{\frac23}(\alpha+2\mu)                  &0                                              &\frac43\sqrt{\frac{10}{3}}(\alpha-\mu)\\
-\frac{4\sqrt{10}}{9}(\beta+2\nu)                    &\frac43\sqrt{\frac{10}{3}}(\alpha-\mu)         &-\frac{8\sqrt{2}}{9}(\beta-\nu)\\
\frac{4\sqrt{2}}{3}\beta                            &-4\sqrt{\frac23}\alpha                            &\frac{4\sqrt{10}}{3}\beta
\end{pmatrix},
\end{gather}
\begin{gather}
Z_{J=\frac32}=\mathrm{diag}\left(\frac{8}{3}(C_{12}-4C_{13}+2C_{45}),\frac83(C_{12}+2C_{13})-16C_{45},\frac83(C_{12}+2C_{13}+2C_{45})\right);
\end{gather}
\begin{gather}
Y_{J=\frac12}=\begin{pmatrix}
0                                                   &\frac43\sqrt{\frac23}(2\alpha+\mu)             &-\frac{8}{3\sqrt{3}}(\alpha+2\mu)\\
\frac43\sqrt{\frac23}(2\alpha+\mu)                  &-\frac{8\sqrt{2}}{9}(2\beta+\nu)              &-\frac89(\beta+2\nu)\\
-\frac{8}{3\sqrt{3}}(\alpha+2\mu)                    &-\frac89(\beta+2\nu)                         &-\frac{20\sqrt{2}}{9}(\beta-\nu)\\
0                                                    &4\sqrt{\frac23}\alpha                        &\frac{8}{\sqrt{3}}\alpha\\
4\sqrt{\frac23}\alpha                                &-\frac{8\sqrt{2}}{3}\beta                    &\frac83\beta
\end{pmatrix},
\end{gather}
\begin{gather}
Z_{J=\frac12}=\mathrm{diag}\Big(\frac83(C_{12}-4C_{13})-16C_{45},\frac83(C_{12}-4C_{13}+2C_{45}),\frac83(C_{12}+2C_{13}+2C_{45})\Big),
\end{gather}
where $\alpha=C_{14}+C_{15}$, $\beta=C_{14}-C_{15}$, $\mu=C_{34}+C_{35}$, and $\nu=C_{34}-C_{35}$. For the $I=0,Y=0$ case, the $Y$ and $Z$ blocks are
\begin{gather}
Y_{J=\frac32}=\begin{pmatrix}
\frac{4\sqrt{2}}{3}\beta\\
-4\sqrt{\frac23}\alpha\\
\frac{4\sqrt{10}}{3}\beta\\
-\frac{4\sqrt{2}}{3}\nu
\end{pmatrix},
\end{gather}
\begin{gather}
Z_{J=\frac32}=-8C_{12}+\frac{16}{3}C_{45};
\end{gather}
\begin{gather}
Y_{J=\frac12}=\begin{pmatrix}
0                           &4\sqrt{\frac23}\alpha\\
4\sqrt{\frac23}\alpha       &-\frac{8\sqrt{2}}{3}\beta\\
\frac{8}{\sqrt{3}}\alpha    &\frac83\beta \\
0                           &-4\sqrt{\frac23}\mu\\
-4\sqrt{\frac23}\mu         &\frac{8\sqrt{2}}{3}\nu\\
\end{pmatrix},
\end{gather}
\begin{gather}
Z_{J=\frac12}=\mathrm{diag}\left(-8(C_{12}+2C_{45}),-8C_{12}+\frac{16}{3}C_{45}\right),
\end{gather}
For the $I=\frac12$, $Y=-1$ ($I=0$, $Y=-2$) case, the matrices are similar to the $I=1$, $Y=0$  $(I=\frac32$, $Y=1)$ case.

\subsection{Rearrangement decay}

In previous works \cite{Cheng:2019obk,Cheng:2020nho}, a simple decay scheme with a constant Hamiltonian $H=\alpha$ has been adopted in order to study the rearrangement decay properties of a multiquark state into two conventional hadrons. In principle, the decay constant $\alpha$ should be changed for different systems. From our study, one finds that the theoretical ratios between widths of $P^N_{\psi}(4312)^+$, $P^N_{\psi}(4440)^+$, and $P^N_{\psi}(4457)^+$ by using this simple model are roughly consistent with the experimental results. Here, we still adopt this model to investigate decay properties of the hidden-charm pentaquark states.

There are four possible rearrangement decay types,
\begin{eqnarray}
(q_1q_2q_3)(c\bar{c})\to(q_1q_2c)_{1c}+(q_3\bar{c})_{1c},\notag \\
(q_1q_2q_3)(c\bar{c})\to(q_1cq_3)_{1c}+(q_2\bar{c})_{1c}, \\
(q_1q_2q_3)(c\bar{c})\to(cq_2q_3)_{1c}+(q_1\bar{c})_{1c},\notag \\
(q_1q_2q_3)(c\bar{c})\to(q_1q_2q_3)_{1c}+(c\bar{c})_{1c}.\notag
\end{eqnarray}
To calculate their matrix elements, one projects the wave function of the final meson-baryon state onto the initial pentaquark. In the color space, the final state is recoupled to the $(qqq)(c\bar{c})$ base by using the SU(3) Clebsch-Gordan coefficients \cite{deSwart:1963pdg,Kaeding:1995vq},
\begin{equation}
\begin{split}
(q_1q_2c)_1(q_3\bar{c})_1&=\frac{2\sqrt{2}}{3}(q_1q_2q_3)_{MA}(c\bar{c})_8+\frac{1}{3}(q_1q_2q_3)_{1}(c\bar{c})_1,\\
(q_1cq_3)_1(q_2\bar{c})_1&=-\sqrt{\frac{2}{3}}(q_1q_2q_3)_{MS}(c\bar{c})_8-\frac{\sqrt{2}}{3}(q_1q_2q_3)_{MA}(c\bar{c})_8+\frac13(q_1q_2q_3)_{1}(c\bar{c})_1,\\
(cq_2q_3)_1(q_1\bar{c})_1&=\sqrt{\frac{2}{3}}(q_1q_2q_3)_{MS}(c\bar{c})_8-\frac{\sqrt{2}}{3}(q_1q_2q_3)_{MA}(c\bar{c})_8+\frac13(q_1q_2q_3)_{1}(c\bar{c})_1,\\
(q_1q_2q_3)_1(c\bar{c})_1&=(q_1q_2q_3)_{1}(c\bar{c})_1.
\end{split}
\end{equation}
In the spin and flavor spaces, similar recouplings are also conducted. The initial wave function of a pentaquark state, as an eigenstate of the chromomagnetic interaction, can be written as $\Psi_{penta}=\sum_{i}x_{i}(q_1q_2q_3c\bar{c})$ where $x_i$ is the element of an eigenvactor of the CMI matrix. Then the amplitude squared of a rearrangement decay channel is $|\mathcal{M}|^2=\alpha^2|\sum_{i}x_{i}y_{i}|^2$. Here, $y_{i}$ represents the coefficient when one recouples the meson-baryon base to the $(qqq)(c\bar{c})$ base. The rearrangement decay width for a pentaquark is then given by
\begin{eqnarray}
\Gamma=|\mathcal{M}|^2\frac{|\vec{p}_1|}{8\pi M^2_{pentaquark}}.
\end{eqnarray}
where $\vec{p}_1$ is the three momentum of a final hadron in the center-of-mass frame.

%%%%%%%%%%%%%%%%%%%%%%%%%%%%%%%%%%%%%%%%%%%
\section{Numerical results}\label{sec3}
%%%%%%%%%%%%%%%%%%%%%%%%%%%%%%%%%%%%%%%%%%%

\subsection{Model parameters}

In our calculations, we use the coupling parameters listed in the last column of Table \ref{effectiveparameters2}. They are extracted from the experimental masses of the conventional ground hadrons. We show the adopted hadrons and related CMI formulas in the first four columns of Table \ref{effectiveparameters2}. More information about the extraction procedure is given in Ref. \cite{Wu:2018xdi}. We will set $m_n=362$ MeV, $m_s=540$ MeV, $m_c=1725$ MeV, and $m_b=5053$ MeV \cite{Wu:2018xdi} for the effective quark masses when adopting Eq. \eqref{mass1}. The mass gap $\Delta_{sn}=90.6$ MeV extracted from ground hadrons is taken from Ref. \cite{Cheng:2020nho}. One may consult Ref. \cite{Cheng:2020nho} for details regarding the selection procedure for this parameter. The masses of final hadrons used in calculations are taken from the particle data book \cite{ParticleDataGroup:2022pth}. Here, we assume that the two-body rearrangement decays saturate the total width.
That is, the sum of two-body rearrangement decay widths is equal to the measured width for a hidden-charm pentaquark state, $\Gamma_{sum}=\Gamma_{total}$. One determines the parameter $\alpha=4647.94$ MeV from the decay width of $P_c(4312)^+$.

\begin{table}[!htb]
	\caption{Chromomagnetic interactions for various hadrons and  obtained effective coupling parameters $C_{ij}$'s in units of MeV.}\setlength{\tabcolsep}{1.3mm}\label{effectiveparameters2}
	\centering
	\begin{tabular}{ccccc}
		\hline\hline
  	Hadron      &$\langle H_{CMI}\rangle$          &Hadron   &$\langle H_{CMI}\rangle$&$C_{ij}$ \\\hline
  	  $N$       &$-8C_{nn}$                        &$\Delta$ &$8C_{nn}$               &$C_{nn}=18.3$     \\
  	  $\Sigma$  &$\frac83C_{nn}-\frac{32}{3}C_{ns}$&$\Sigma^*$ &$\frac83C_{nn}+\frac{16}{3}C_{ns}$   &$C_{ns}=12.0$     \\
  	  $\Sigma_c$&$\frac83C_{nn}-\frac{32}{3}C_{cn}$&$\Sigma_c^*$ &$\frac83C_{nn}+\frac{16}{3}C_{cn}$ &$C_{cn}=4.0$  \\
  	  $\Xi_c^{\prime}$ &$\frac83C_{ns}-\frac{16}{3}C_{cn}-\frac{16}{3}C_{cs}$&$\Xi_c^*$&$\frac83C_{ns}+\frac{8}{3}C_{cn}+\frac{8}{3}C_{cs}$&$C_{cs}=4.4$\\
  	  $\eta_c$&$-16C_{c\bar{c}}$&$J/\psi$&$\frac{16}{3}C_{c\bar{c}}$&$C_{c\bar{c}}=5.3$\\
  	  $D_s$&$-16C_{c\bar{s}}$&$D^*_s$&$\frac{16}{3}C_{c\bar{s}}$&$C_{c\bar{s}}=6.7$ \\
  	  $D$&$-16C_{c\bar{n}}$&$D^*$&$\frac{16}{3}C_{c\bar{n}}$&$C_{c\bar{n}}=6.6$ \\
  	  $\Omega$&$8C_{ss}$&&&$C_{ss}=5.7$\\
  	  \hline \hline
	\end{tabular}\\
\end{table}

With the above parameters, the masses and decay widths of ground hidden-charm pentaquark states are calculated. We list these results in Tables \ref{nnncc-mass}-\ref{decay5}.

\subsection{The $nnnc\bar{c}$ system}\label{sec.3.2}

There are four $I(J^P)=\frac12(\frac32^-)$ $nnnc\bar{c}$ states when one considers contributions from both color-octet and color-singlet structures. Following the conclusion of Ref. \cite{Cheng:2019obk}, we assume that the $P^N_{\psi}(4312)^+$ is the second lowest $I(J^P)=\frac12(\frac32^-)$ $nnnc\bar{c}$ compact pentaquark and treat it as the reference state in studying other pentaquarks.

We collect the numerical results for the masses of $nnnc\bar{c}$ compact states in Table \ref{nnncc-mass}. In the table, the first column shows the quantum numbers. The second and third columns list the numerical values for the CMI matrix and the corresponding eigenvalues, respectively. The fourth column gives the pentaquark masses by referencing to $P^N_{\psi}(4312)^+$. The masses in the fifth and sixth columns are estimated with the $NJ/\psi$ ($\Delta J/\psi$) threshold and Eq. \eqref{mass1}, respectively. They can be treated as the lower and upper limits for the masses of the $nnnc\bar{c}$ states.

\begin{table}[htbp]\centering
\caption{Calculated CMI eigenvalues and estimated masses for the $nnnc\bar{c}$ pentaquark states in units of MeV. The masses in the forth, fifth, and sixth columns are obtained with $P^N_{\psi}(4312)^+$, meson-baryon threshold, and effective quark masses, respectively.}  \scriptsize\label{nnncc-mass}
	\begin{tabular}{c|cccccc}\hline
		\hline
		$I(J^{P})$ & $\langle H_{CMI} \rangle$ &Eigenvalue & Mass& $J/\Psi N$ ($J/\psi \Delta$) &Upper limits\\
		$\frac{1}{2}(\frac{5}{2}^-)$ &$\left(\begin{array}{c}96.7\end{array}\right)$&$\left(\begin{array}{c}96.7\end{array}\right)$&$\left(\begin{array}{c}4479.2\end{array}\right)$&$\left(\begin{array}{c}4250.7\end{array}\right)$&$\left(\begin{array}{c}4810.9\end{array}\right)$\\
		$\frac{1}{2}(\frac{3}{2}^-)$ &$\left(\begin{array}{cccc}-18.9&49.6&47.2&-6.9\\49.6&47.2&-20.1&34.6\\47.2&-20.1&-9.3&11.0\\-6.9&34.6&11.0&-118.1\end{array}\right)$&$\left(\begin{array}{c}78.0\\26.8\\-70.7\\-133.3\end{array}\right)$&$\left(\begin{array}{c}4460.6\\4409.3\\4311.9\\4249.3\end{array}\right)$&$\left(\begin{array}{c}4232.0\\4180.8\\4083.3\\4020.7\end{array}\right)$&$\left(\begin{array}{c}4792.2\\4741.0\\4643.5\\4580.9\end{array}\right)$\\
		$\frac{1}{2}(\frac{1}{2}^-)$ &$\left(\begin{array}{ccccc}-26.0&-9.0&-70.2&0&49.0\\-9.0&-82.5&29.9&49.0&13.9\\-70.2&29.9&-72.9&-49.0&6.9\\0&49.0&-49.0&-231.2&0\\49.0&13.9&6.9&0&-118.1\end{array}\right)$&$\left(\begin{array}{c}38.2\\-58.6\\-91.3\\-155.4\\-263.6\end{array}\right)$&$\left(\begin{array}{c}4420.7\\4323.9\\4291.2\\4227.2\\4118.9\end{array}\right)$&$\left(\begin{array}{c}4192.2\\4095.4\\4062.7\\3998.6\\3890.4\end{array}\right)$&$\left(\begin{array}{c}4752.4\\4655.6\\4622.9\\4558.8\\4450.6\end{array}\right)$\\
		\hline
%		$I(J^{P})$ & $\langle H_{CMI} \rangle$ &Eigenvalue & Mass& $J/\Psi\Delta $ &Upper limits\\
		$\frac{3}{2}(\frac{5}{2}^-)$ &$\left(\begin{array}{c}174.7\end{array}\right)$&$\left(\begin{array}{c}174.7\end{array}\right)$&$\left(\begin{array}{c}4557.2\end{array}\right)$&$\left(\begin{array}{c}4329.5\end{array}\right)$&$\left(\begin{array}{c}4888.9\end{array}\right)$\\
		$\frac{3}{2}(\frac{3}{2}^-)$ &$\left(\begin{array}{ccc}170.8&49.0&15.5\\49.0&61.6&0\\15.5&0&174.7\end{array}\right)$&$\left(\begin{array}{c}198.4\\166.0\\42.6\end{array}\right)$&$\left(\begin{array}{c}4581.0\\4548.6\\4425.2\end{array}\right)$&$\left(\begin{array}{c}4353.3\\4320.8\\4197.5\end{array}\right)$&$\left(\begin{array}{c}4912.6\\4880.2\\4756.8\end{array}\right)$\\
		$\frac{3}{2}(\frac{1}{2}^-)$ &$\left(\begin{array}{ccc}193.6&61.2&-69.2\\61.2&196.8&9.8\\-69.2&9.8&174.7\end{array}\right)$&$\left(\begin{array}{c}277.5\\196.8\\90.7\end{array}\right)$&$\left(\begin{array}{c}4660.1\\4579.3\\4473.3\end{array}\right)$&$\left(\begin{array}{c}4432.4\\4351.6\\4245.6\end{array}\right)$&$\left(\begin{array}{c}4991.7\\4911.0\\4804.9\end{array}\right)$\\
		\hline\hline
	\end{tabular}
\end{table}

\begin{figure}[htbp]\centering
\begin{tabular}{ccc}
	\includegraphics[width=180pt]{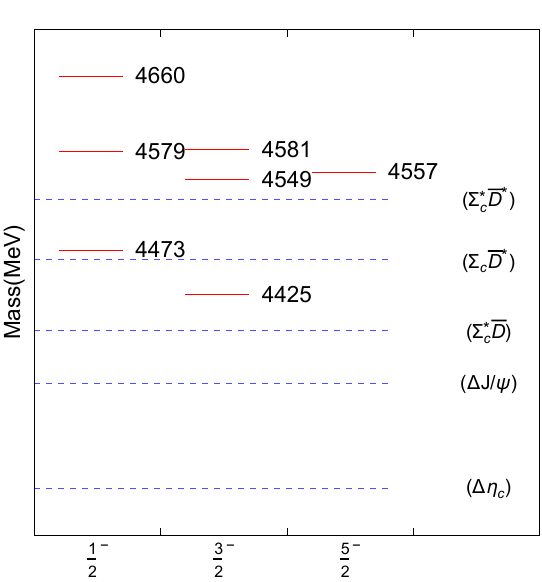}&$\qquad$&\includegraphics[width=180pt]{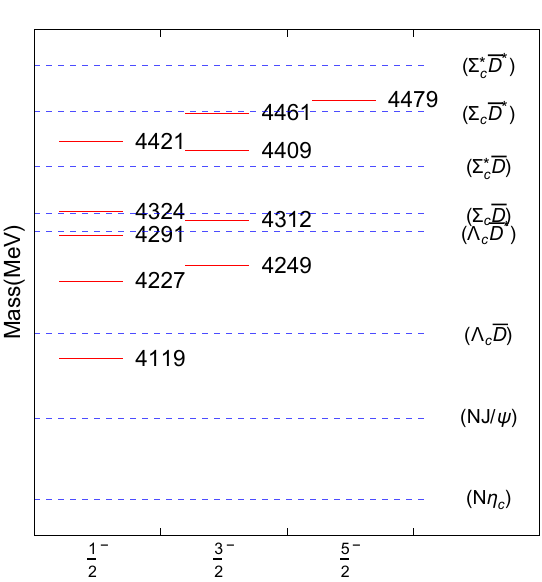}\\
	(a) $I=\frac32$ $nnnc\bar{c}$ states & $\qquad$& (b) $I=\frac12$ $nnnc\bar{c}$ states
\end{tabular}
\caption{Relative positions for the hidden-charm pentaquark states. The red solid and blue dashed lines correspond to the pentaquark masses and various thresholds, respectively.}\label{nnncc-picture}
\end{figure}

\begin{table}[htbp]\centering
	\caption{Rearrangement decay widths for the $I=\frac{1}{2}, Y=1$ $nnnc\bar{c}$ states in units of MeV.}\label{decay2}
	\begin{tabular}{c|ccccccccc}\hline\hline
		$I(J^P)=\frac12(\frac52^-)$&$\Sigma_c^{*}\bar{D}^{*}$&&&&&&&&$\Gamma_{sum}$\\
		4479.2&(11.1,$-$)&&&&&&&&0.0\\
		\hline
		$I(J^P)=\frac12(\frac32^-)$&$\Sigma_c^{*}\bar{D^*}$&$\Sigma_c^{*}\bar{D}$&$\Sigma_c \bar{D^*}$&$\Lambda_c \bar{D^*}$&$N J/\psi$ &&&&$\Gamma_{sum}$\\
		4460.6&(32.7,$-$)&(3.0,1.6)&(4.6,$-$)&(3.5,2.8)&(2.0,0.7)&&&&5.0\\
		4409.3&(1.3,$-$)&(1.3,0.4)&(36.0,$-$)&(5.8,3.8)&(0.2,0.1)&&&&4.2\\
		4311.9&(0.0,$-$)&(20.2,$-$)&(0.9,$-$)&(17.7,4.6)&(17.1,5.2)&&&&9.8\\
		4249.3&(1.2,$-$)&(14.5,$-$)&(1.1,$-$)&(0.8,$-$)&(80.7,22.0)&&&&22.0\\
		\hline
		$I(J^P)=\frac12(\frac12^-)$&$\Sigma_c^{*}\bar{D^*}$&$\Sigma_c \bar{D^*}$&$\Sigma_c \bar{D}$&$\Lambda_{c}\bar{D^*}$ &$\Lambda_{c}\bar{D}$ &$N J/\psi$ &$N \eta_{c}$ &&$\Gamma_{sum}$\\
		4420.7&(18.8,$-$)&(9.1,$-$)&(1.1,0.6)&(13.3,9.1)&(0.6,0.6)&(4.4,1.5)&(0.6,0.2)&&12.1\\
		4323.9&(7.9,$-$)&(20.5,$-$)&(0.8,0.1)&(0.7,0.2)&(8.2,6.8)&(16.9,5.2)&(2.3,0.9)&&13.2\\
		4291.2&(2.3,$-$)&(0.9,$-$)&(15.4,$-$)&(12.1,$-$)&(2.6,2.0)&(19.4,5.7)&(14.0,5.0)&&12.7\\
		4227.2&(0.1,$-$)&(0.2,$-$)&(10.5,$-$)&(1.0,$-$)&(12.3,6.9)&(59.3,15.4)&(1.7,0.6)&&22.9\\
		4118.9&(0.5,$-$)&(0.9,$-$)&(11.2,$-$)&(0.7,$-$)&(4.1,$-$)&(0.0,0.0)&(81.4,22.5)&&22.5\\
		\hline\hline
	\end{tabular}
\end{table}

\begin{table}[htbp]\centering
	\caption{Rearrangement decay widths for the $I=\frac{3}{2}, Y=1$ $nnnc\bar{c}$ states in units of MeV.}\label{decay1}
	\begin{tabular}{c|ccccccccc}\hline\hline
		$I(J^P)=\frac32(\frac52^-)$&$\Sigma_c^{*}\bar{D}^{*}$&$\Delta J/\psi$&&&&&&&$\Gamma_{sum}$\\
		4557.2&(11.1,3.6)&(100.0,26.9)&&&&&&&30.4\\
		\hline
		$I(J^P)=\frac32(\frac32^-)$&$\Sigma_c^{*}\bar{D^*}$&$\Sigma_c^{*}\bar{D}$&$\Sigma_c \bar{D^*}$&$\Delta J/\psi$ &$\Delta \eta_c$  &&&&$\Gamma_{sum}$\\
		4581.0&(24.0,10.2)&(2.2,1.8)&(6.4,4.1)&(8.2,2.3)&(27.4,9.3)&&&&27.6\\
		4548.6&(5.6,1.5)&(10.8,8.0)&(2.2,1.2)&(5.0,1.3)&(72.4,23.7)&&&&35.8\\
		4425.2&(0.1,$-$)&(9.2,3.5)&(6.2,$-$)&(86.8,15.8)&(0.2,0.0)&&&&19.4\\
		\hline
		$I(J^P)=\frac32(\frac12^-)$&$\Sigma_c^{*}\bar{D^*}$&$\Sigma_c \bar{D^*}$&$\Sigma_c \bar{D}$&$\Delta J/\psi$ &&&&&$\Gamma_{sum}$\\
		4660.1&(37.2,24.3)&(0.5,0.4)&(0.1,0.1)&(20.1,6.3)&&&&&31.0\\
		4579.3&(1.0,0.4)&(28.8,18.1)&(0.0,0.0)&(43.7,12.2)&&&&&30.8\\
		4473.3&(2.5,$-$)&(7.8,1.5)&(22.2,16.3)&(36.2,7.9)&&&&&25.8\\
		\hline\hline
	\end{tabular}
\end{table}

Fig. \ref{nnncc-picture} displays the relative positions for the $nnnc\bar{c}$ compact states. In the $I=1/2$ case, four pentaquark states are located above 4.4 GeV and three pentaquarks have masses around 4.3 GeV. The results indicate that one may identify the calculated $J^P=\frac32^-$ $(J^P=\frac12^-)$ pentaquark with mass 4461 (4421) MeV to be the $P^N_{\psi}(4457)^+$ ($P^N_{\psi}(4440)^+$), which is consistent with the assignment given in Ref. \cite{Cheng:2019obk}. Just from the mass, the $P^N_{\psi}(4337)^+$ seems to be a $J=\frac12$ state. One can check whether this assignment is reasonable from the decay properties.

\begin{figure}[htbp]
	\centering
	\includegraphics[width=200pt]{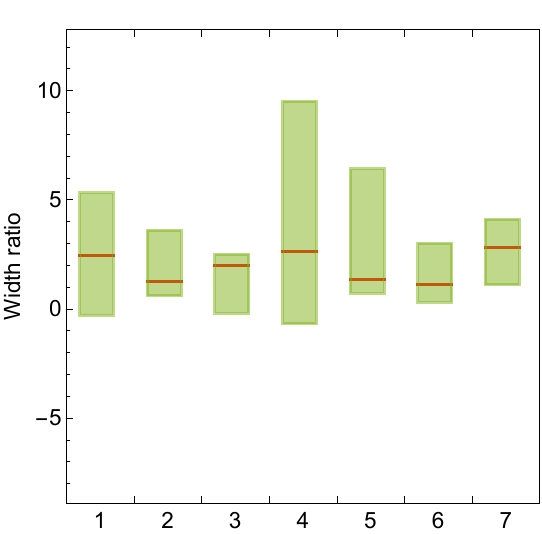}
	\caption{Ratios between decay widths of different pentaquarks. Those obtained from the model calculation and experimental data are represented by red solid lines and green rectangles, respectively. The numbers 1, 2, $\cdots$, 7 below the $x$-axis correspond to $\frac{\Gamma(P^N_{\psi}(4440)^+)}{\Gamma(P^N_{\psi}(4457)^+)}$, $\frac{\Gamma(P^N_{\psi}(4440)^+)}{\Gamma(P^N_{\psi}(4312)^+)}$, $\frac{\Gamma(P^N_{\psi}(4312)^+)}{\Gamma(P^N_{\psi}(4457)^+)}$, $\frac{\Gamma(P^N_{\psi}(4337)^+)}{\Gamma(P^N_{\psi}(4457)^+)}$, $\frac{\Gamma(P^N_{\psi}(4337)^+)}{\Gamma(P^N_{\psi}(4312)^+)}$, $\frac{\Gamma(P^N_{\psi}(4337)^+)}{\Gamma(P^N_{\psi}(4440)^+)}$, and $\frac{\Gamma(P^{\Lambda}_{\psi s}(4459)^0)}{\Gamma(P^{\Lambda}_{\psi s}(4338)^0)}$, respectively. \label{decay ratio}}.
\end{figure}

In Table \ref{decay2} (Table \ref{decay1}), we present the rearrangement decay widths for the $I=1/2$ ($I=3/2$) $nnnc\bar{c}$ pentaquarks. The ratios between widths of the interested states will be checked. To avoid confusion, we use the symbol $\tilde{P}$ to denote theoretical states. From the results in Table \ref{decay2}, one gets
\begin{eqnarray}\label{Theory}
\Gamma(\tilde{P}^N_{\psi}(4421)^+):\Gamma(\tilde{P}^N_{\psi}(4461)^+)=2.42 \notag, \\
\Gamma(\tilde{P}^N_{\psi}(4421)^+):\Gamma(\tilde{P}^N_{\psi}(4312)^+)=1.24 \notag, \\
\Gamma(\tilde{P}^N_{\psi}(4312)^+):\Gamma(\tilde{P}^N_{\psi}(4461)^+)=1.96,\\
\Gamma(\tilde{P}^N_{\psi}(4324)^+):\Gamma(\tilde{P}^N_{\psi}(4461)^+)=2.64\notag,\\
\Gamma(\tilde{P}^N_{\psi}(4324)^+):\Gamma(\tilde{P}^N_{\psi}(4312)^+)=1.35\notag,\\
\Gamma(\tilde{P}^N_{\psi}(4324)^+):\Gamma(\tilde{P}^N_{\psi}(4421)^+)=1.09.\notag
\end{eqnarray}
On the other hand, the ratios between the measured widths are
\begin{eqnarray}\label{Exp1}
\Gamma(P^N_{\psi}(4440)^+):\Gamma(P^N_{\psi}(4457)^+)=3.2^{+2.1}_{-3.5} \notag,\\
\Gamma(P^N_{\psi}(4440)^+):\Gamma(P^N_{\psi}(4312)^+)=2.1^{+1.5}_{-1.5} \notag,\\
\Gamma(P^N_{\psi}(4312)^+):\Gamma(P^N_{\psi}(4457)^+)=1.5^{+1.0}_{-1.7}, \\
\Gamma(P^N_{\psi}(4337)^+):\Gamma(P^N_{\psi}(4457)^+)=4.5^{+5.0}_{-5.2}\notag,\\
\Gamma(P^N_{\psi}(4337)^+):\Gamma(P^N_{\psi}(4312)^+)=3.0^{+3.4}_{-2.3}\notag,\\
\Gamma(P^N_{\psi}(4337)^+):\Gamma(P^N_{\psi}(4440)^+)=1.4^{+1.6}_{-1.1}.\notag
\end{eqnarray}
In order to clearly compare the results in Eq. \eqref{Theory} with those in Eq. \eqref{Exp1}, we plot the values of ratios in Fig. \ref{decay ratio}. One finds that the calculated ratios between widths are compatible with the experimental data within error. Then it is reasonable to regard the $P^N_{\psi}(4457)^+$, $P^N_{\psi}(4440)^+$, and $P^N_{\psi}(4337)^+$ as our $\tilde{P}^N_{\psi}(4461)$ with $I(J^P)=\frac12(\frac32^-)$, $\tilde{P}^N_{\psi}(4421)$ with $I(J^P)=\frac12(\frac12^-)$, and $\tilde{P}^N_{\psi}(4324)$ with $I(J^P)=\frac12(\frac12^-)$, respectively.

If the above assignment is correct, we can give an estimate for the partial width ratios for the four $P^N_{\psi}$ states. In the $P^N_{\psi}(4457)^+$ case, one has $\Gamma(\Sigma^*_c \bar{D}):\Gamma(\Lambda_c \bar{D}^*):\Gamma(N J/\Psi)=2.3:4.0:1.0$. Since the contributions from the color-singlet component are included now, the hidden-charm decay modes can be described. The $P^N_{\psi}(4440)^+$ would mainly decay into $\Lambda_c\bar{D}^*$, while its decays into $\Sigma_c\bar{D}$, $\Lambda_c\bar{D}$, $N J/\Psi$, and $N\eta_c$ are relatively suppressed. The ratios between partial widths of these five channels are $45.5:3.0:3.0:7.5:1.0$. For the $P^N_{\psi}(4312)^+$, the partial width ratio between the two dominant decay modes $\Lambda_c\bar{D}^*$ and $N J/\Psi$ is $\Gamma(N J/\Psi):\Gamma(\Lambda_c\bar{D}^*)=1.1$. This is different from our previous result \cite{Cheng:2019obk}. The $P^N_{\psi}(4337)^+$ may have two dominant decay channels $\Lambda_c\bar{D}$ and $N J/\Psi$ with the branching fraction reaching up to $91\%$. The ratio between their partial widths is found to be   $\Gamma(\Lambda_c\bar{D}):\Gamma(NJ/\Psi)=1.3$.
 The decay into $N\eta_c$ is also sizable with the branching fraction of Br$[P^N_{\psi}(4337)^+\to N\eta_c]\sim7\%$. However, the decay channels $\Sigma_c\bar{D}$ and $\Lambda_c\bar{D}^*$ are suppressed. If our results are all acceptable, it is worth noting that the $I(J^P)=\frac12(\frac52^-)$ hidden-charm state $\tilde{P}^N_{\psi}(4479)^+$, a compact structure without hidden-charm decay channels, may be stable, because its mass is lower than the $\Sigma_c^*\bar{D}^*$ threshold. Beside these five states, four additional pentaquarks may also exist whose decay properties can be found in Table \ref{decay2}.

Compared with the $I=\frac12$ $nnnc\bar{c}$ pentaquarks, the masses and rearrangement decay widths of the $I=\frac32$ states are overall larger. All the $I=\frac32$ states can decay into $\Delta J/\psi$ and search for all of them in this mode is possible. However, the $J=\frac52$ state ($\tilde{P}^N_{\psi}(4557)$), the two heaviest $J=\frac32$ states ($\tilde{P}^N_{\psi}(4581)$ and $\tilde{P}^N_{\psi}(4549)$), and the second heaviest $J=\frac12$ state ($\tilde{P}^N_{\psi}(4579)$) have similar masses, which probably makes it difficult to distinguish them in a preliminary experimental study. The $\tilde{P}^N_{\psi}(4557)$ mainly decays into $\Delta J/\psi$ and $\Sigma_c^*\bar{D}^*$, while the $J=\frac32$ ($J=\frac12$) states have special rearrangement channels $\Sigma_c^*\bar{D}$ and $\Delta \eta_c$ ($\Sigma_c\bar{D}$).

The above discussions are based on the assignment that the $P^N_{\psi}(4312)^+$ is a compact pentaquark with $I(J^P)=\frac12(\frac32^-)$. This assumption results from the combined analysis of mass spectrum and decay properties. To see the consistency between the present study and the study in Ref. \cite{Cheng:2019obk}, we list the eigenvalues and eigenvectors of the $I(J^P)=\frac12(\frac32^-)$ CMI matrix in Table \ref{ratio3/2}. Clearly, the color-octet component dominates the wave function of $P^N_{\psi}(4312)^+$ with a probability $\sim83\%$.

\begin{table}[!htb]
	\caption{Ratios between the color-octet and color-singlet components of $I(J^P)=\frac12(\frac32^-)$ $nnnc\bar{c}$ compact states}\label{ratio3/2}
	\centering
	\begin{tabular}{|c|c|c|cccccccccccccc}
		\hline
		Eigenvalue (MeV)&Eigenvector&Ratio\\\hline
		78.0&\{0.264, -0.270, -0.224, 0.898\}&0.980:0.020\\
		26.8&\{0.489, -0.324, 0.809, -0.040\}&0.998:0.002\\
		-70.7&\{-0.686, 0.259, 0.539, 0.414\}&0.829:0.171\\
		-133.3&\{0.264, -0.270, -0.224, 0.898\}&0.193:0.807\\
		\hline
	\end{tabular}\\
\end{table}

\subsection{The $nnsc\bar{c}$ system}

\setlength{\tabcolsep}{1.3mm}
\begin{table}[htbp]\centering
	\caption{Calculated CMI eigenvalues and estimated masses for the $nnsc\bar{c}$ pentaquark states in units of MeV. The masses in the forth, fifth, and sixth columns are obtained with $P^N_{\psi}(4312)^+$, meson-baryon threshold, and effective quark masses, respectively.}\scriptsize\label{nnscc-mass}
	\begin{tabular}{c|cccccc}\hline
		\hline
		$I(J^{P})$ & $\langle H_{CMI} \rangle$ &Eigenvalue & Mass& $J/\Psi \Sigma$ ($J/\psi\Lambda$) &Upper limits\\\hline
		$1(\frac{5}{2}^-)$ &$\left(\begin{array}{cc}101.9&-0.6\\-0.6&141.1\end{array}\right)$&$\left(\begin{array}{c}141.1\\101.9\end{array}\right)$&$\left(\begin{array}{c}4614.2\\4575.0\end{array}\right)$&$\left(\begin{array}{c}4478.7\\4439.5\end{array}\right)$&$\left(\begin{array}{c}4855.3\\4816.1\end{array}\right)$\\
		$1(\frac{3}{2}^-)$ &$\left(\begin{array}{ccccccc}67.2&35.0&34.9&-57.1&-4.7&35.7&10.1\\35.0&51.4&-20.1&35.1&35.7&0&-1.2\\34.9&-20.1&-5.8&33.4&10.1&-1.2&0.4\\-57.1&35.1&33.4&76.9&-4.9&-34.6&-11.0\\-4.7&35.7&10.1&-4.9&-50.9&0&0\\35.7&0&-1.2&-34.6&0&28.0&0\\10.1&-1.2&0.4&-11.0&0&0&141.1\end{array}\right)$&$\left(\begin{array}{c}159.9\\131.1\\92.2\\39.4\\7.5\\-36.5\\-85.7\end{array}\right)$&$\left(\begin{array}{c}4633.1\\4604.2\\4565.4\\4512.5\\4480.7\\4436.6\\4387.5\end{array}\right)$&$\left(\begin{array}{c}4497.6\\4468.7\\4429.9\\4377.0\\4345.2\\4301.1\\4252.0\end{array}\right)$&$\left(\begin{array}{c}4874.1\\4845.3\\4806.4\\4753.6\\4721.7\\4677.7\\4628.5\end{array}\right)$\\
		$1(\frac{1}{2}^-)$ &$\left(\begin{array}{cccccccc}75.4&26.1&-49.5&-72.0&-35.1&0&35.2&-50.5\\26.1&49.4&22.1&-35.1&-101.9&35.2&9.4&6.4\\-49.5&22.1&-70.4&-49.6&21.1&-50.5&6.4&0.9\\-72.0&-35.1&-49.6&83.8&28.1&0&34.6&49.0\\-35.1&-101.9&21.1&28.1&55.1&34.6&9.8&-6.9\\0&35.2&-50.5&0&34.6&-164.0&0&0\\35.2&9.4&6.4&34.6&9.8&0&-50.9&0\\-50.5&6.4&0.9&49.0&-6.9&0&0&141.1\end{array}\right)$&$\left(\begin{array}{c}238.9\\157.8\\66.7\\49.9\\-19.9\\-52.9\\-110.5\\-210.4\end{array}\right)$&$\left(\begin{array}{c}4712.0\\4630.9\\4539.9\\4523.0\\4453.3\\4420.2\\4362.6\\4262.8\end{array}\right)$&$\left(\begin{array}{c}4576.5\\4495.4\\4404.4\\4387.5\\4317.7\\4284.7\\4227.1\\4127.2\end{array}\right)$&$\left(\begin{array}{c}4953.1\\4872.0\\4780.9\\4764.1\\4694.3\\4661.3\\4603.7\\4503.8\end{array}\right)$\\
		\hline
%		$I(J^{P})$ & $\langle H_{CMI} \rangle$ &Eigenvalue & Mass& $J/\Psi\Lambda $ &Upper limits\\\hline
		$0(\frac{5}{2}^-)$ &$\left(\begin{array}{c}76.7\end{array}\right)$&$\left(\begin{array}{c}76.7\end{array}\right)$&$\left(\begin{array}{c}4549.8\end{array}\right)$&$\left(\begin{array}{c}4407.4\end{array}\right)$&$\left(\begin{array}{c}4790.9\end{array}\right)$\\
		$0(\frac{3}{2}^-)$ &$\left(\begin{array}{ccccc}-72.1&-36.4&-31.9&-57.1&-4.9\\-36.4&26.2&-18.7&35.1&-34.6\\-31.9&-18.7&-31.0&33.4&-11.0\\-57.1&35.1&33.4&-112.8&4.3\\-4.9&-34.6&-11.0&4.3&-118.1\end{array}\right)$&$\left(\begin{array}{c}60.0\\5.0\\-82.0\\-135.3\\-155.5\end{array}\right)$&$\left(\begin{array}{c}4533.1\\4478.2\\4391.2\\4337.9\\4317.6\end{array}\right)$&$\left(\begin{array}{c}4390.7\\4335.7\\4248.7\\4195.4\\4175.2\end{array}\right)$&$\left(\begin{array}{c}4774.2\\4719.2\\4632.2\\4578.9\\4558.7\end{array}\right)$\\
		$0(\frac{1}{2}^-)$ &$\left(\begin{array}{ccccccc}-93.8&-44.8&51.5&-72.0&-35.1&0&34.6\\-44.8&-179.5&-20.2&-35.1&-101.9&34.6&9.8\\51.5&-20.2&-95.6&-49.6&21.1&49.0&-6.9\\-72.0&-35.1&-49.6&-135.8&-44.0&0&-36.3\\-35.1&-101.9&21.1&-44.0&-224.2&-36.3&-8.7\\0&34.6&49.0&0&-36.3&-231.2&0\\34.6&9.8&-6.9&-36.3&-8.7&0&-118.1\end{array}\right)$&$\left(\begin{array}{c}23.8\\-71.8\\-101.8\\-145.5\\-168.9\\-268.0\\-346.2\end{array}\right)$&$\left(\begin{array}{c}4497.0\\4401.4\\4371.4\\4327.7\\4304.3\\4205.1\\4127.0\end{array}\right)$&$\left(\begin{array}{c}4354.5\\4258.9\\4229.0\\4185.2\\4161.9\\4062.7\\3984.5\end{array}\right)$&$\left(\begin{array}{c}4738.0\\4642.4\\4612.4\\4568.7\\4545.3\\4446.2\\4368.0\end{array}\right)$\\
		\hline\hline
	\end{tabular}
\end{table}

\begin{figure}[htbp]\centering
	\begin{tabular}{ccc}
		\includegraphics[width=180pt]{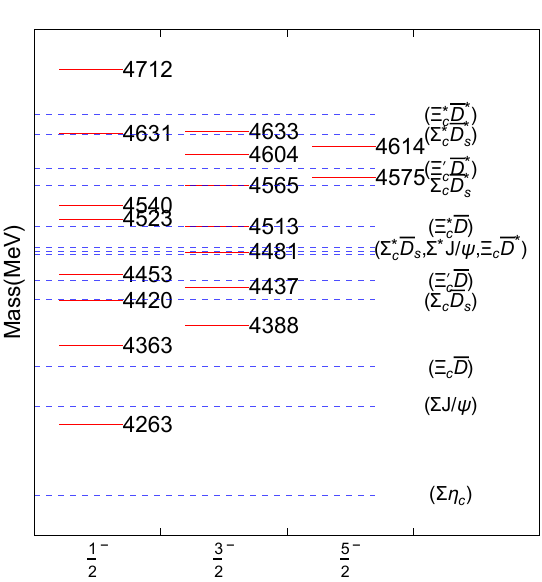}&$\qquad$&
		\includegraphics[width=180pt]{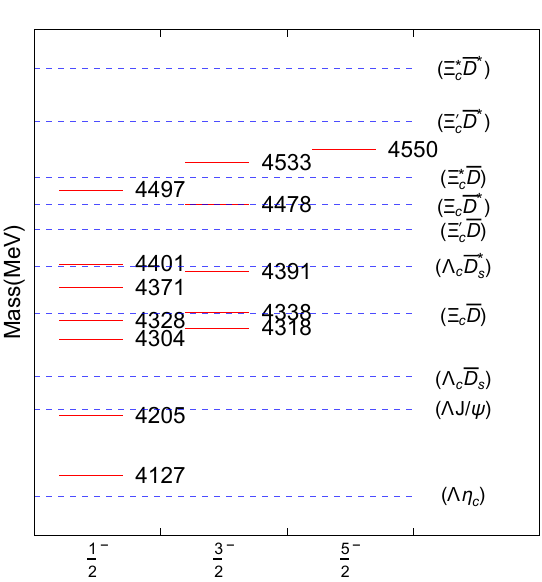}\\
		(a) $I=1$ $nnsc\bar{c}$ states & $\qquad$& (b) $I=0$ $nnsc\bar{c}$ states
	\end{tabular}
	\caption{Relative positions for the hidden-charm pentaquark states. The red solid and blue dashed lines correspond to the pentaquark masses and various thresholds, respectively.}\label{nnscc-picture}
\end{figure}

The masses of the $nnsc\bar{c}$ compact pentaquarks are calculated and are listed in Table \ref{nnscc-mass}. We depict the relative positions for these states in Fig. \ref{nnscc-picture}. In the $I=0$ case, five pentaquarks have masses around 4338 MeV and two pentaquarks have masses close to 4459 MeV. Just from the spectrum, the theoretical $\tilde{P}^{\Lambda}_{\psi s}(4338)$ and $\tilde{P}^{\Lambda}_{\psi s}(4478)$ with $J=\frac32$ are good candidates for the $P^{\Lambda}_{\psi s}(4338)^0$ and $P^{\Lambda}_{\psi s}(4459)^0$, respectively, but there are also other possibilities. To discuss possible assignments for the quantum numbers of the two observed $P^{\Lambda}_{\psi s}$ states, we again adopt the decay widths estimated with the simple rearrangement scheme. The results in the isoscalar case are summarized in Table \ref{decay4}.

If one assigns the $P^{\Lambda}_{\psi s}(4459)^0$ and $P^{\Lambda}_{\psi s}(4338)^0$ to be $J=\frac32$ pentaquark states $\tilde{P}^{\Lambda}_{\psi s}(4478)^0$ and $\tilde{P}^{\Lambda}_{\psi s}(4338)^0$, respectively, the calculated width ratio $\Gamma(\tilde{P}^{\Lambda}_{\psi s}(4478)^0):\Gamma(\tilde{P}^{\Lambda}_{\psi s}(4338)^0)=0.12$ is contradicted with the experimental value $\Gamma(P^{\Lambda}_{\psi s}(4459)^0):\Gamma(P^{\Lambda}_{\psi s}(4338)^0)=2.5^{+1.6}_{-1.4}$. We have to consider other possible assignments. The relevant width ratios are
\begin{eqnarray}
\Gamma(\tilde{P}^{\Lambda}_{\psi s}(4478)^0):\Gamma(\tilde{P}^{\Lambda}_{\psi s}(4371)^0)&=&0.15,\nonumber\\
\Gamma(\tilde{P}^{\Lambda}_{\psi s}(4478)^0):\Gamma(\tilde{P}^{\Lambda}_{\psi s}(4328)^0)&=&0.56,\nonumber\\
\Gamma(\tilde{P}^{\Lambda}_{\psi s}(4478)^0):\Gamma(\tilde{P}^{\Lambda}_{\psi s}(4318)^0)&=&2.57,\nonumber\\
\Gamma(\tilde{P}^{\Lambda}_{\psi s}(4478)^0):\Gamma(\tilde{P}^{\Lambda}_{\psi s}(4304)^0)&=&0.17,\nonumber\\
\Gamma(\tilde{P}^{\Lambda}_{\psi s}(4497)^0):\Gamma(\tilde{P}^{\Lambda}_{\psi s}(4371)^0)&=&0.72,\nonumber\\
\Gamma(\tilde{P}^{\Lambda}_{\psi s}(4497)^0):\Gamma(\tilde{P}^{\Lambda}_{\psi s}(4338)^0)&=&0.61,\nonumber\\
\Gamma(\tilde{P}^{\Lambda}_{\psi s}(4497)^0):\Gamma(\tilde{P}^{\Lambda}_{\psi s}(4328)^0)&=&2.78,\nonumber\\
\Gamma(\tilde{P}^{\Lambda}_{\psi s}(4497)^0):\Gamma(\tilde{P}^{\Lambda}_{\psi s}(4318)^0)&=&12.71,\nonumber\\
\Gamma(\tilde{P}^{\Lambda}_{\psi s}(4497)^0):\Gamma(\tilde{P}^{\Lambda}_{\psi s}(4304)^0)&=&0.83.
\end{eqnarray}
The third and seventh ratios are consistent with the experimental value. However, the width of $\tilde{P}^{\Lambda}_{\psi s}(4318)^0$ is much smaller than the measured $\Gamma(P^{\Lambda}_{\psi s}(4338)^0)$, which leads to the most possible assignment that the observed $P^{\Lambda}_{\psi s}(4459)^0$ and $P^{\Lambda}_{\psi s}(4338)^0$ correspond to $\tilde{P}^{\Lambda}_{\psi s}(4497)^0$ and $\tilde{P}^{\Lambda}_{\psi s}(4328)^0$, respectively. Therefore, our analysis indicates that the quantum numbers of both $P^{\Lambda}_{\psi s}(4338)^0$ and $P^{\Lambda}_{\psi s}(4459)^0$ may be assigned as $I(J^P)=0(\frac12^-)$. The comparison of width ratio between model calculation and experimental value with this assignment is also shown in Fig. \ref{decay ratio}.

If the $P^{\Lambda}_{\psi s}(4459)^0$ indeed corresponds to the highest $J=\frac12$ pentaquark state $\tilde{P}^{\Lambda}_{\psi s}(4497)^0$, it may mainly decay into $\Lambda_c\bar{D}^*_s$, $\Xi_c\bar{D}^*$, and $\Lambda J/\Psi$, while the decays into $\Lambda_c\bar{D}_s$, $\Xi_c^{\prime}\bar{D}$, $\Xi_c \bar{D}$, and $\Lambda\eta_c$  are suppressed because of  small phase space. The ratios between the main partial widths of $P^{\Lambda}_{\psi s}(4459)^0$ are predicted to be $\Gamma(\Lambda_c\bar{D}^*_s):\Gamma(\Xi_c\bar{D}^*):\Gamma(\Lambda J/\Psi)=2.3:1.1:1.0$, which can be tested in future experiments. If the $P^{\Lambda}_{\psi s}(4338)^0$ really corresponds to the fourth highest $J=\frac12$ pentaquark state $\tilde{P}^{\Lambda}_{\psi s}(4328)^0$, its main decay modes would be $\Lambda J/\Psi$ and $\Lambda_c\bar{D}_s$. The ratio between the corresponding partial widths is estimated to be $\Gamma(\Lambda J/\Psi):\Gamma(\Lambda_c\bar{D}_s)=3.0$.

It is interesting to note that the $J=\frac52$ state $\tilde{P}^{\Lambda}_{\psi s}(4550)^0$, the lightest $J=\frac32$ state $\tilde{P}^{\Lambda}_{\psi s}(4318)^0$, and the lightest $J=\frac12$ state $\tilde{P}^{\Lambda}_{\psi s}(4127)^0$ may be stable. The $\tilde{P}^{\Lambda}_{\psi s}(4550)^0$ being a compact hidden-color structure can be searched for in the radiative decay channel $\Xi_c^{*+}D^-\gamma$. The search for $\tilde{P}^{\Lambda}_{\psi s}(4318)^0$ can be conducted with more analyses in the $\Lambda J/\Psi$ channel. The experimentalists may search for the $\tilde{P}^{\Lambda}_{\psi s}(4127)^0$ in the $\Lambda^0 \eta_c$ or $\Lambda^0 \pi^+D_s^-$ channel.

In the $I=1$ case, many $nnsc\bar{c}$ states have the $\Sigma^* J/\psi$ decay mode. They can be searched for in this channel. Of course, other channels listed in Table \ref{decay3} can also be used. The light $J=\frac52$ pentaquark state $\tilde{P}^{\Lambda}_{\psi s}(4575)$ should be a stable one, which can be searched for in the $\Sigma^{*++} J/\Psi$ and $\Lambda_c^+\pi^- D^{*-}_s$ channels.

\begin{table}[htbp]\centering
	\caption{Rearrangement decay widths for the $I=0, Y=0$ $nnsc\bar{c}$ states in units of MeV. }\label{decay4}
	\begin{tabular}{c|cccccccccc}\hline\hline
		$I(J^P)=0(\frac52^-)$&$\Xi_c^{*}\bar{D}^{*}$&&&&&&&&&$\Gamma_{sum}$\\
		{{4549.8}}&(66.7,$-$)&&&&&&&&&0.0\\
		\hline
		$I(J^P)=0(\frac32^-)$&$\Lambda_c \bar{D_s^*}$&$\Xi_c^* \bar{D^*}$&$\Xi_c^* \bar{D}$&$\Xi_c^\prime \bar{D^*}$&$\Xi_c \bar{D^*}$&$\Lambda J/\psi$&&&&$\Gamma_{sum}$\\
		4533.1&(6.0,1.4)&(49.9,$-$)&(4.2,0.7)&(5.4,$-$)&(3.0,0.9)&(2.3,0.7)&&&&3.7\\
		4478.2&(9.3,1.7)&(1.2,$-$)&(1.3,$-$)&(54.9,$-$)&(4.4,$-$)&(0.3,0.1)&&&&1.8\\
		4391.2&(20.0,$-$)&(0.1,$-$)&(24.8,$-$)&(2.1,$-$)&(17.5,$-$)&(24.3,6.0)&&&&6.0\\
		4337.9&(11.0,$-$)&(1.0,$-$)&(21.7,$-$)&(1.3,$-$)&(2.6,$-$)&(69.3,14.6)&&&&14.6\\
		4317.6&(53.7,$-$)&(0.6,$-$)&(6.3,$-$)&(0.2,$-$)&(30.9,$-$)&(3.8,0.7)&&&&0.7\\
		\hline
		$I(J^P)=0(\frac12^-)$&$\Lambda_c \bar{D_s^*}$&$\Lambda_c \bar{D_s}$&$\Xi_c^* \bar{D^*}$&$\Xi_c^\prime \bar{D^*}$&$\Xi_c^\prime \bar{D}$&$\Xi_c \bar{D^*}$&$\Xi_c \bar{D}$&$\Lambda J/\psi$&$\Lambda \eta_{c}$&$\Gamma_{sum}$\\
		4497.0&(21.0,4.2)&(0.9,0.3)&(28.3,$-$)&(12.1,$-$)&(1.2,0.3)&(11.6,2.0)&(0.4,0.2)&(5.8,1.8)&(0.6,0.2)&8.9\\
		4401.4&(0.8,0.0)&(10.6,2.6)&(10.8,$-$)&(31.2,$-$)&(0.6,$-$)&(1.0,$-$)&(4.9,1.6)&(20.5,5.2)&(3.0,1.0)&10.4\\
		4371.4&(12.1,$-$)&(6.0,1.3)&(4.4,$-$)&(2.3,$-$)&(18.4,$-$)&(12.5,$-$)&(2.0,0.5)&(25.1,5.9)&(15.1,4.7)&12.4\\
		4327.7&(40.7,$-$)&(4.2,0.8)&(0.1,$-$)&(0.2,$-$)&(2.4,$-$)&(30.0,$-$)&(3.1,$-$)&(12.1,2.4)&(0.0,0.0)&3.2\\
		4304.3&(24.9,$-$)&(18.9,2.8)&(0.2,$-$)&(0.0,$-$)&(16.1,$-$)&(3.9,$-$)&(5.4,$-$)&(36.4,6.6)&(4.5,1.2)&10.7\\
		4205.1&(0.4,$-$)&(4.9,$-$)&(0.6,$-$)&(1.2,$-$)&(19.5,$-$)&(0.5,$-$)&(3.9,$-$)&(0.1,$-$)&(76.5,15.6)&15.6\\
		4127.0&(0.1,$-$)&(54.6,$-$)&(0.1,$-$)&(0.1,$-$)&(0.1,$-$)&(0.1,$-$)&(38.7,$-$)&(0.0,$-$)&(0.3,0.0)&0.0\\
		\hline\hline
	\end{tabular}
\end{table}

\setlength{\tabcolsep}{1.0mm}
\begin{table}[htbp]\centering\scriptsize
	\caption{Rearrangement decay widths for the $I=1, Y=0$ $nnsc\bar{c}$ states in units of MeV.}\label{decay3}
	\begin{tabular}{c|cccccccccccc}\hline\hline
		$I(J^P)=1(\frac52^-)$&$\Sigma_c^{*}\bar{D}_s^{*}$&$\Xi_c^{*}\bar{D}^{*}$&$\Sigma^{*}J/\psi$&&&&&&&&&$\Gamma_{sum}$\\
		4614.2&(10.2,$-$)&(11.6,$-$)&(100.0,20.6)&&&&&&&&&20.6\\
		4575.0&(89.8,$-$)&(21.8,$-$)&(0.0,0.0)&&&&&&&&&0.0\\
		\hline
		$I(J^P)=1(\frac32^-)$&$\Sigma_c^{*}\bar{D_s^*}$&$\Sigma_c^{*}\bar{D_s}$&$\Sigma_c \bar{D_s^*}$&$\Xi_c^* \bar{D^*}$&$\Xi_c^* \bar{D}$&$\Xi_c^\prime \bar{D^*}$&$\Xi_c \bar{D^*}$&$\Sigma^{*}J/\psi$&$\Sigma^{*}\eta_{c}$&$\Sigma J/\psi$&&$\Gamma_{sum}$\\
		4633.1&(18.8,0.6)&(1.3,0.3)&(6.9,1.1)&(23.4,$-$)&(1.3,0.5)&(6.7,1.8)&(0.0,0.0)&(35.5,7.8)&(8.0,2.3)&(0.0,0.0)&&14.4\\
		4604.2&(7.3,$-$)&(11.1,2.3)&(2.0,0.2)&(8.0,$-$)&(11.4,4.1)&(1.5,0.3)&(0.0,0.0)&(64.3,12.8)&(6.7,1.9)&(0.0,0.0)&&21.6\\
		4565.4&(72.4,$-$)&(3.9,0.7)&(3.1,$-$)&(15.5,$-$)&(1.1,0.3)&(0.9,$-$)&(7.9,2.9)&(0.0,0.0)&(0.1,0.0)&(3.2,0.9)&&4.8\\
		4512.5&(0.3,$-$)&(0.2,0.0)&(72.1,$-$)&(0.0,$-$)&(0.1,$-$)&(18.2,$-$)&(11.0,2.6)&(0.0,0.0)&(0.0,0.0)&(2.0,0.5)&&3.1\\
		4480.7&(0.0,$-$)&(7.5,$-$)&(7.1,$-$)&(0.0,$-$)&(10.7,$-$)&(5.9,$-$)&(0.0,0.0)&(0.2,$-$)&(85.0,16.9)&(0.0,0.0)&&17.0\\
		4436.6&(0.6,$-$)&(15.0,$-$)&(8.3,$-$)&(0.2,$-$)&(2.8,$-$)&(2.7,$-$)&(22.0,$-$)&(0.0,$-$)&(0.1,0.0)&(53.9,12.2)&&12.2\\
		4387.5&(0.6,$-$)&(61.1,$-$)&(0.5,$-$)&(0.1,$-$)&(14.3,$-$)&(0.1,$-$)&(0.7,$-$)&(0.0,$-$)&(0.0,0.0)&(40.8,7.7)&&7.7\\
		\hline
		$I(J^P)=1(\frac12^-)$&$\Sigma_c^{*}\bar{D_s^*}$&$\Sigma_c\bar{D_s^*}$&$\Sigma_c \bar{D_s}$&$\Xi_c^* \bar{D^*}$&$\Xi_c^\prime \bar{D^*}$&$\Xi_c^\prime \bar{D}$&$\Xi_c \bar{D^*}$&$\Xi_c \bar{D}$&$\Sigma^{*}J/\psi$&$\Sigma\eta_{c}$&$\Sigma J/\psi$&$\Gamma_{sum}$\\
		4712.0&(35.2,5.9)&(0.3,0.1)&(0.0,0.0)&(37.1,10.4)&(0.3,0.1)&(0.0,0.0)&(0.0,0.0)&(0.0,0.0)&(23.0,6.0)&(0.0,0.0)&(0.0,0.0)&22.6\\
		4630.9&(1.3,0.0)&(27.9,4.3)&(0.0,0.0)&(1.8,$-$)&(28.1,7.1)&(0.0,0.0)&(0.0,0.0)&(0.0,0.0)&(44.1,9.6)&(0.0,0.0)&(0.0,0.0)&21.1\\
		4539.9&(36.0,$-$)&(8.3,$-$)&(2.1,0.5)&(10.6,$-$)&(5.0,$-$)&(0.1,0.0)&(21.4,6.6)&(0.9,0.5)&(0.7,0.1)&(0.6,0.2)&(9.2,2.6)&10.5\\
		4523.0&(6.6,$-$)&(11.6,$-$)&(20.0,4.0)&(1.3,$-$)&(7.1,$-$)&(22.4,7.8)&(0.5,0.1)&(0.0,0.0)&(32.2,3.8)&(0.0,0.0)&(0.3,0.1)&15.8\\
		4453.3&(9.8,$-$)&(47.3,$-$)&(0.4,0.1)&(2.4,$-$)&(11.1,$-$)&(0.1,0.0)&(2.8,$-$)&(6.9,3.0)&(0.0,$-$)&(2.9,0.9)&(25.5,6.0)&10.0\\
		4420.2&(8.2,$-$)&(2.5,$-$)&(12.8,$-$)&(2.3,$-$)&(0.7,$-$)&(3.2,$-$)&(12.0,$-$)&(9.3,3.5)&(0.0,$-$)&(18.8,5.5)&(34.4,7.4)&16.3\\
		4362.6&(0.2,$-$)&(1.0,$-$)&(30.0,$-$)&(0.0,$-$)&(0.3,$-$)&(7.1,$-$)&(4.7,$-$)&(19.0,4.0)&(0.0,$-$)&(9.6,2.5)&(30.5,5.0)&11.5\\
		4262.8&(0.0,$-$)&(1.1,$-$)&(34.5,$-$)&(0.1,$-$)&(0.2,$-$)&(8.5,$-$)&(0.2,$-$)&(5.5,$-$)&(0.0,$-$)&(68.1,12.7)&(0.1,$-$)&12.7\\
		\hline\hline
	\end{tabular}
\end{table}

\subsection{The $ssnc\bar{c}$ system}

\begin{table}[htbp]\centering
	\caption{Calculated CMI eigenvalues and estimated masses for the $ssnc\bar{c}$ pentaquark states in units of MeV. The masses in the forth, fifth, and sixth columns are obtained with $P^N_{\psi}(4312)^+$, meson-baryon threshold, and effective quark masses, respectively.}\scriptsize\label{ssncc-mass}
	\begin{tabular}{c|cccccc}\hline
		\hline\hline
		$I(J^{P})$ & $\langle H_{CMI} \rangle$ &Eigenvalue & Mass& $J/\Psi \Xi$ &Upper limits\\\hline
		$0(\frac{5}{2}^-)$ &$\left(\begin{array}{cc}69.3&0.6\\0.6&107.5\end{array}\right)$&$\left(\begin{array}{c}107.5\\69.3\end{array}\right)$&$\left(\begin{array}{c}4671.2\\4633.0\end{array}\right)$&$\left(\begin{array}{c}4607.2\\4569.0\end{array}\right)$&$\left(\begin{array}{c}5002.3\\4964.1\end{array}\right)$\\
		$0(\frac{3}{2}^-)$ &$\left(\begin{array}{ccccccc}35.3&36.1&31.9&-57.1&-4.5&35.2&10.5\\36.1&17.8&-17.9&36.0&35.2&0&1.2\\31.9&-17.9&-40.1&33.4&10.5&1.2&-0.4\\-57.1&36.0&33.4&25.5&-4.3&-36.3&-9.7\\-4.5&35.2&10.5&-4.3&-84.5&0&0\\35.2&0&1.2&-36.3&0&-5.6&0\\10.5&1.2&-0.4&-9.7&0&0&107.5\end{array}\right)$&$\left(\begin{array}{c}121.9\\95.6\\57.4\\-0.4\\-27.9\\-71.1\\-119.7\end{array}\right)$&$\left(\begin{array}{c}4685.7\\4659.3\\4621.2\\4563.3\\4535.9\\4492.7\\4444.0\end{array}\right)$&$\left(\begin{array}{c}4621.6\\4595.3\\4557.2\\4499.3\\4471.9\\4428.7\\4380.0\end{array}\right)$&$\left(\begin{array}{c}5016.7\\4990.4\\4952.2\\4894.4\\4866.9\\4823.7\\4775.1\end{array}\right)$\\
		$0(\frac{1}{2}^-)$ &$\left(\begin{array}{cccccccc}41.8&28.1&-51.1&-72.0&-36.0&0&35.7&-49.7\\28.1&12.5&20.2&-36.0&-101.9&35.7&9.1&6.7\\-51.1&20.2&-105.7&-50.9&21.1&-49.7&6.7&-0.9\\-72.0&-36.0&-50.9&33.4&26.1&0&36.3&51.3\\-36.0&-101.9&21.1&26.1&6.7&36.3&8.7&-6.1\\0&35.7&-49.7&0&36.3&-197.6&0&0\\35.7&9.1&6.7&36.3&8.7&0&-84.5&0\\-49.7&6.7&-0.9&51.3&-6.1&0&0&107.5\end{array}\right)$&$\left(\begin{array}{c}200.4\\118.7\\29.2\\9.2\\-60.1\\-87.4\\-150.4\\-245.6\end{array}\right)$&$\left(\begin{array}{c}4764.2\\4682.5\\4593.0\\4573.0\\4503.7\\4476.4\\4413.3\\4318.1\end{array}\right)$&$\left(\begin{array}{c}4700.1\\4618.5\\4529.0\\4508.9\\4439.6\\4412.3\\4349.3\\4254.1\end{array}\right)$&$\left(\begin{array}{c}5095.2\\5013.5\\4924.0\\4904.0\\4834.7\\4807.4\\4744.4\\4649.2\end{array}\right)$\\
		\hline\hline
	\end{tabular}
\end{table}

\begin{figure}[htbp]\centering
	\begin{tabular}{ccc}
		\includegraphics[width=180pt]{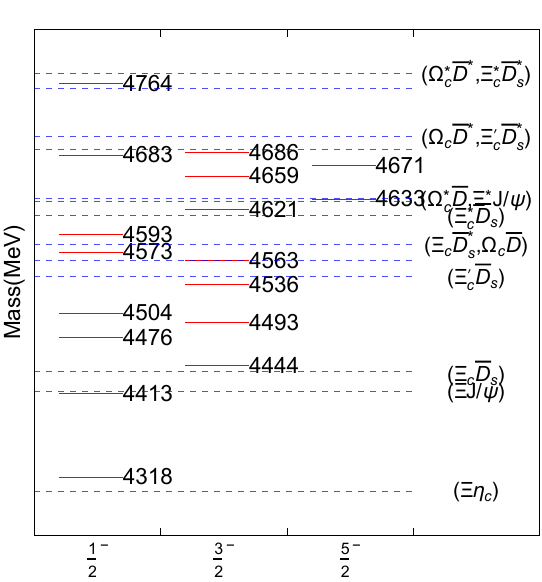}&$\qquad$&
		\includegraphics[width=180pt]{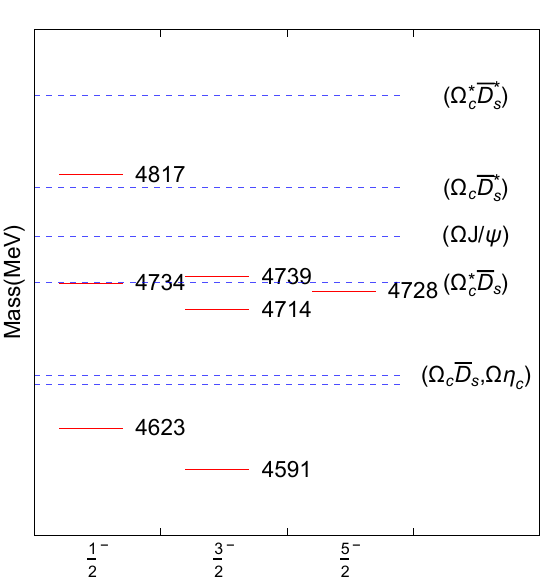}\\
		(a) $I=\frac12$ $ssnc\bar{c}$ states & $\qquad$& (b) $I=0$ $sssc\bar{c}$ states
	\end{tabular}
	\caption{Relative positions for the hidden-charm pentaquark states. The red solid and blue dashed lines correspond to the pentaquark masses and various thresholds, respectively.}\label{ssscc-picture}
\end{figure}

\setlength{\tabcolsep}{1.3mm}
\begin{table}[htbp]\centering\scriptsize
\caption{Rearrangement decay widths for the $I=\frac{1}{2}, Y=-1$ $ssnc\bar{c}$ states in units of MeV.}\label{decay6}
	\begin{tabular}{c|cccccccccccc}\hline \hline
		$I(J^P)=\frac12(\frac52^-)$&$\Omega_c^{*}\bar{D}^{*}$&$\Xi_c^{*}\bar{D}_s^{*}$&$\Xi^{*}J/\psi$&&&&&&&&&$\Gamma_{sum}$\\
		4671.2&(12.1,$-$)&(10.6,$-$)&(100.0,11.5)&&&&&&&&&11.5\\
		4633.0&(87.9,$-$)&(22.7,$-$)&(0.0,0.0)&&&&&&&&&0.0\\
		\hline
		$I(J^P)=\frac12(\frac32^-)$&$\Omega_c^{*}\bar{D^*}$&$\Omega_c^{*}\bar{D}$&$\Omega_c \bar{D^*}$&$\Xi_c^* \bar{D_s^*}$&$\Xi_c^* \bar{D_s}$&$\Xi_c^\prime \bar{D_s^*}$&$\Xi_c \bar{D_s^*}$&$\Xi^{*}J/\psi$&$\Xi^{*}\eta_{c}$&$\Xi J/\psi$&&$\Gamma_{sum}$\\
		4685.7&(21.8,$-$)&(0.5,0.1)&(7.0,$-$)&(17.5,$-$)&(0.5,0.2)&(7.3,$-$)&(0.0,0.0)&(46.0,6.1)&(7.3,1.7)&(0.0,0.0)&&8.0\\
		4659.3&(11.5,$-$)&(11.9,1.1)&(0.9,$-$)&(10.3,$-$)&(11.5,2.9)&(1.2,$-$)&(0.0,0.0)&(53.7,5.2)&(9.2,2.0)&(0.0,0.0)&&11.2\\
		4621.2&(65.0,$-$)&(4.7,$-$)&(4.5,$-$)&(19.0,$-$)&(1.1,0.1)&(1.0,$-$)&(6.8,1.7)&(0.0,$-$)&(0.1,0.0)&(3.4,0.9)&&2.6\\
		4563.3&(0.4,$-$)&(0.7,$-$)&(72.9,$-$)&(0.2,$-$)&(0.1,$-$)&(18.0,$-$)&(10.5,$-$)&(0.0,$-$)&(0.0,0.0)&(1.5,0.3)&&0.3\\
		4535.9&(0.0,$-$)&(12.5,$-$)&(5.8,$-$)&(0.0,$-$)&(8.9,$-$)&(6.9,$-$)&(0.0,$-$)&(0.2,$-$)&(83.2,6.8)&(0.1,0.0)&&6.8\\
		4492.7&(0.6,$-$)&(11.0,$-$)&(8.5,$-$)&(0.1,$-$)&(3.7,$-$)&(1.6,$-$)&(23.2,$-$)&(0.0,$-$)&(0.1,$-$)&(56.0,9.1)&&9.1\\
		4444.0&(0.6,$-$)&(58.7,$-$)&(0.5,$-$)&(0.2,$-$)&(15.8,$-$)&(0.2,$-$)&(1.0,$-$)&(0.0,$-$)&(0.0,$-$)&(39.0,3.9)&&3.9\\
		\hline
		$I(J^P)=\frac12(\frac12^-)$&$\Omega_c^{*}\bar{D^*}$&$\Omega_c\bar{D^*}$&$\Omega_c \bar{D}$&$\Xi_c^* \bar{D_s^*}$&$\Xi_c^\prime \bar{D_s^*}$&$\Xi_c^\prime \bar{D_s}$&$\Xi_c \bar{D_s^*}$&$\Xi_c \bar{D_s}$&$\Xi^{*}J/\psi$&$\Xi\eta_{c}$&$\Xi J/\psi$&$\Gamma_{sum}$\\
		4764.2&(36.8,$-$)&(0.1,0.0)&(0.0,0.0)&(34.9,3.2)&(0.1,0.0)&(0.0,0.0)&(0.0,0.0)&(0.0,0.0)&(26.1,5.2)&(0.0,0.0)&(0.0,0.0)&8.5\\
		4682.5&(2.9,$-$)&(27.2,$-$)&(0.1,0.0)&(2.2,$-$)&(27.2,$-$)&(0.1,0.1)&(0.0,0.0)&(0.0,0.0)&(44.0,5.7)&(0.0,0.0)&(0.0,0.0)&5.7\\
		4593.0&(39.6,$-$)&(15.0,$-$)&(0.1,0.0)&(9.0,$-$)&(1.5,$-$)&(1.0,0.3)&(21.1,2.8)&(0.8,0.4)&(0.4,$-$)&(0.6,0.2)&(10.3,2.4)&6.1\\
		4573.0&(0.7,$-$)&(7.2,$-$)&(23.4,1.4)&(4.2,$-$)&(10.7,$-$)&(21.0,4.3)&(0.3,$-$)&(0.0,0.0)&(29.4,$-$)&(0.0,0.0)&(0.3,0.1)&5.7\\
		4503.7&(7.7,$-$)&(46.3,$-$)&(0.8,$-$)&(2.2,$-$)&(12.3,$-$)&(0.2,$-$)&(2.4,$-$)&(5.0,1.6)&(0.0,$-$)&(2.6,0.7)&(31.2,5.4)&7.7\\
		4476.4&(9.1,$-$)&(1.4,$-$)&(12.2,$-$)&(2.8,$-$)&(0.3,$-$)&(3.0,$-$)&(12.8,$-$)&(10.2,2.6)&(0.0,$-$)&(20.5,5.0)&(30.2,4.4)&12.0\\
		4413.3&(0.4,$-$)&(1.5,$-$)&(28.6,$-$)&(0.1,$-$)&(0.3,$-$)&(7.6,$-$)&(4.9,$-$)&(20.0,$-$)&(0.0,$-$)&(9.4,1.9)&(28.0,$-$)&1.9\\
		4318.1&(0.0,$-$)&(1.2,$-$)&(34.8,$-$)&(0.1,$-$)&(0.3,$-$)&(8.8,$-$)&(0.1,$-$)&(5.7,$-$)&(0.0,$-$)&(66.8,5.3)&(0.1,$-$)&5.3\\
		\hline \hline
	\end{tabular}
\end{table}

The symmetry for the wave functions of $ssnc\bar{c}$ states is the same as that for $I=1$, $Y=0$ $nnsc\bar{c}$ states. Noticing the difference in effective coupling parameters, one can get similar CMI matrices from those for $nnsc\bar{c}$. The numerical results are collected in Table \ref{ssncc-mass} where the data listed in the fourth, fifth, and sixth columns are estimated with the $P^N_{\psi}(4312)^+$, $J/\Psi\Xi$ threshold, and effective quark masses, respectively. We also plot the relative positions for pentaquark states and relevant meson-baryon thresholds in Fig. \ref{ssscc-picture}(a). The rearrangement decay information can be found from Table \ref{decay6}.

From the results, the lightest state whose spin is 1/2 has a mass around $4.3$ GeV. It has only one rearrangement decay channel $\Xi\eta_c$. Although the coupling with this channel is strong, the width is not large because of the small phase space. The rearrangement decay width of the light $J=\frac52$ pentaquark is very small, which indicates that it is also stable. Searching for such a state in the $\Xi^*J/\psi$ channel will give more information. The fourth highest $J=\frac32$ state also has a relatively stable structure. It may be searched for in the $\Xi^*\eta_c$ and $\Xi J/\Psi$ channels. Compared with the $nnnc\bar{c}$ and $nnsc\bar{c}$ cases, the rearrangement decay widths in the $ssnc\bar{c}$ case are relatively smaller. It is possible to observe many double-strange hidden-charm exotic structures in the $\Xi J/\psi$ or $\Xi^* J/\psi$ mass distribution. The open-charm decay channels listed in Table \ref{decay6} may be used to distinguish the spins of the observed structures.

\subsection{The $sssc\bar{c}$ system}

As for the $sssc\bar{c}$ case, the calculation procedure and resulting expressions are similar to the $I=\frac32,Y=1$ $nnnc\bar{c}$ case, but the numerical results are different. We present the mass results in Table \ref{ssscc-mass}, show the relative positions for pentaquarks and relevant meson-baryon thresholds in Fig. \ref{ssscc-picture}(b), and give the rearrangement decay information in Table \ref{decay5}.

From Tables \ref{decay1} and \ref{decay5}, compared with the $I=\frac32,Y=1$ $nnnc\bar{c}$ case, the decay widths of $sssc\bar{c}$ states are relatively small because of the smaller phase space. The model calculation tells us that the lightest $J=\frac12$ state with mass 4623 MeV, the lightest $J=\frac32$ state with mass 4591 MeV, and the $J=\frac52$ state with mass 4728 MeV are below their rearrangement decay thresholds and should all be stable. The search for them in the $\Xi^0\pi^-J/\psi$ channel is called for. The second lightest $J=\frac12$ pentaquark with mass 4734 MeV has one rearrangement decay channel $\Omega_c\bar{D}_s$. Although it is higher than the threshold, the coupling with this channel is weak. It should also be a stable state and a search for this pentaquark in the $\Omega_c\bar{D}_s$ channel is strongly proposed.

\begin{table}[htbp]\centering
	\caption{Calculated CMI eigenvalues and estimated masses for the $sssc\bar{c}$ pentaquark states in units of MeV. The masses in the forth, fifth, and sixth columns are obtained with $P^N_{\psi}(4312)^+$, meson-baryon threshold, and effective quark masses, respectively.}\scriptsize\label{ssscc-mass}
	\begin{tabular}{c|cccccc}\hline\hline
		$I(J^{P})$ & $\langle H_{CM} \rangle$ &Eigenvalue & Mass& $J/\Psi\Omega $ &Upper limits\\\hline
		$0(\frac{5}{2}^-)$ &$\left(\begin{array}{c}73.9\end{array}\right)$&$\left(\begin{array}{c}73.9\end{array}\right)$&$\left(\begin{array}{c}4728.2\end{array}\right)$&$\left(\begin{array}{c}4769.4\end{array}\right)$&$\left(\begin{array}{c}5149.3\end{array}\right)$\\
		$0(\frac{3}{2}^-)$ &$\left(\begin{array}{ccc}45.8&51.3&13.7\\51.3&-39.2&0\\13.7&0&73.9\end{array}\right)$&$\left(\begin{array}{c}84.6\\59.4\\-63.5\end{array}\right)$&$\left(\begin{array}{c}4738.9\\4713.8\\4590.8\end{array}\right)$&$\left(\begin{array}{c}4780.1\\4755.0\\4632.0\end{array}\right)$&$\left(\begin{array}{c}5160.0\\5134.8\\5011.9\end{array}\right)$\\
		$0(\frac{1}{2}^-)$ &$\left(\begin{array}{ccc}67.6&64.1&-72.5\\64.1&68.8&8.7\\-72.5&8.7&73.9\end{array}\right)$&$\left(\begin{array}{c}162.2\\79.6\\-31.5\end{array}\right)$&$\left(\begin{array}{c}4816.6\\4734.0\\4622.8\end{array}\right)$&$\left(\begin{array}{c}4857.7\\4775.1\\4664.0\end{array}\right)$&$\left(\begin{array}{c}5237.6\\5155.0\\5043.9\end{array}\right)$\\
		\hline\hline
	\end{tabular}
\end{table}

\begin{table}[htbp]\centering
	\caption{Rearrangement decay widths for the $I=0, Y=-2$ $sssc\bar{c}$ states in units of MeV.}\label{decay5}
	\begin{tabular}{c|ccccccccc}\hline\hline
		$J=0(\frac52^-)$&$\Omega_c^{*}\bar{D}_s^{*}$&$\Omega J/\psi$&&&&&&&$\Gamma_{sum}$\\
		4728.2&(11.1,$-$)&(100.0,$-$)&&&&&&&0.0\\
		\hline
		$J=0(\frac32^-)$&$\Omega_c^{*}\bar{D_s^*}$&$\Omega_c^{*}\bar{D_s}$&$\Omega_c \bar{D_s^*}$&$\Omega J/\psi$ &$\Omega \eta_c$  &&&&$\Gamma_{sum}$\\
		4738.9&(15.4,$-$)&(0.1,0.0)&(7.5,$-$)&(6.1,$-$)&(58.3,9.4)&&&&9.4\\
		4713.8&(14.2,$-$)&(11.6,$-$)&(0.6,$-$)&(12.4,$-$)&(41.5,5.7)&&&&5.7\\
		4590.8&(0.0,$-$)&(10.5,$-$)&(6.7,$-$)&(81.5,$-$)&(0.2,$-$)&&&&0.0\\
		\hline
		$J=(\frac12^-)$&$\Omega_c^{*}\bar{D_s^*}$&$\Omega_c \bar{D_s^*}$&$\Omega_c \bar{D_s}$&$\Omega J/\psi$ &&&&&$\Gamma_{sum}$\\
		4816.6&(34.5,$-$)&(0.0,0.0)&(0.0,0.0)&(29.5,3.5)&&&&&3.5\\
		4734.0&(3.4,$-$)&(26.4,$-$)&(0.2,0.1)&(43.3,$-$)&&&&&0.1\\
		4622.8&(2.8,$-$)&(10.6,$-$)&(22.0,$-$)&(27.1,$-$)&&&&&0.0\\
		\hline\hline
	\end{tabular}
\end{table}

%%%%%%%%%%%%%%%%%%%%%%%%%%%%%%%%%%%%%%%%%%%
\section{Summary}\label{sec4}
%%%%%%%%%%%%%%%%%%%%%%%%%%%%%%%%%%%%%%%%%%%

In this work, we investigate the mass spectra and two-body rearrangement decays of the S-wave hidden-charm pentaquark states within a mass splitting model. In this model, the $P^N_{\psi}(4312)^+$ is assumed to be a hidden-charm compact pentaquark with $I(J^P)=\frac12(\frac32^-)$ and the properties of other pentaquarks are studied by treating the $P^N_{\psi}(4312)^+$ as the reference state. Both color-octet $(qqq)_{8_c}(c\bar{c})_{8_c}$ ($q=u,d,s$) and color-singlet $(qqq)_{1_c}(c\bar{c})_{1_c}$ components are considered for the wave functions.

From the numerical analyses, one finds that the $P^N_{\psi}(4457)^+$, $ P^N_{\psi}(4440)^+$, and $P^N_{\psi}(4337)^+$ can be regarded as the $I(J^P)=\frac12(\frac32^-)$, $\frac12(\frac12^-)$, and $\frac12(\frac12^-)$ pentaquark states, respectively. The $P^N_{\psi}(4457)^+$ mainly rearranges into $\Sigma^*_c\bar{D}$, $\Lambda_c\bar{D}^*$, and $N/J/\Psi$. The dominant decay channel of $P^N_{\psi}(4440)^+$ is $\Lambda_c\bar{D}^*$. For the rearrangement decay of $P^N_{\psi}(4312)^+$, the $NJ/\Psi$ and $\Lambda_c\bar{D}^*$ channel are of equal importance. The $P^N_{\psi}(4337)^+$ mainly decays into $\Lambda_c\bar{D}$ as well as $NJ/\Psi$. The high spin  pentaquark state $nnnc\bar{c}$ ($n=u,d$) with $I(J^P)=\frac12(\frac52^-)$ has a mass around 4479 MeV, but it should be narrow. This state has only color-octet $(nnn)_{8_c}(c\bar{c})_{8_c}$ component and can be searched for in the $\Lambda^+_c\pi^-D^{*+}$ channel in future experiments.

From the spectrum of $I=0, Y=0$ $nnsc\bar{c}$ pentaquark states, we get good candidates of $P^{\Lambda}_{\psi s}(4338)^0$ and $P^{\Lambda}_{\psi s}(4459)^0$ whose quantum numbers are $I(J^P)=0(\frac32^-)$. However, the ratio between their widths cannot be understood. When a slightly larger uncertainty in mass is allowed, we find that assigning both $P^{\Lambda}_{\psi s}(4338)^0$ and $P^{\Lambda}_{\psi s}(4459)^0$ to be pentaquark states with $I(J^P)=0(\frac12^-)$ can result in a width ratio consistent with the experimental data. The lightest isoscalar pentaquarks with $J=\frac12$, $\frac32$, and $\frac52$ should all be narrow states. This $J=\frac52$ state, similar to the case of $I(J^P)=\frac12(\frac52^-)$ $nnnc\bar{c}$, also has only color-octet component.

According to our results for the $ssnc\bar{c}$ case, the light $J=\frac52$ state and the fourth highest $J=\frac32$ state have narrow widths. For the $sssc\bar{c}$ case, there may be four stable states which are the lightest ones with $J=\frac12,\frac32,\frac52$ and the second lightest one with $J=\frac12$.

%And a broad peak  also can be found in the $\Xi^*J/\psi$ invariant mass spectrum at about $\sqrt{s}=4.68$ GeV that stem from the mixing of the first heaviest $3/2^-$ and the second heaviest $1/2^-$ states. One may find a prominent signal corresponding to the highest $3/2^-$ in the $\Xi^*\eta_c$ channel.

%%%%%%%%%%%%%%%%%%%%%%%%%%%%%%%%
\section*{Acknowledgments}
%%%%%%%%%%%%%%%%%%%%%%%%%%%%%%%%

We would like to thank Dr. Jian-Bo Cheng for useful discussions. This project was supported by the National Natural Science Foundation of China under Grant Nos. 12235008, 12275157, 11775132, and 11905114.

\bibliographystyle{unsrt}
\bibliography{inport1}

\end{document}